\documentstyle[proceedings, psfig]{crckapb}
\def\etal{{\it et al.\ }}
\def\eg{{\it e.g.,\ }}

\def\mnras{{\it Mon. Not. R. Astron. Soc.\ }}
\def\gray{$\gamma$-ray\ }
\def\grays{$\gamma$-rays\ }
\def\apj{{\it Astrophys. J.\ }}
\def\prl{{\it Phys. Rev. Letters\ }}

\def\aa{{\it Astron. Astrophys.}}
\def\gray{{$\gamma$-ray}}
\def\grays{{$\gamma$-rays}}
\def\mic{{$\mu$m}}
\def\xxvicrc{{\it 25th International Cosmic Ray Conf.}}

\def\xxviicrc{{\it 26th International Cosmic Ray Conf.}}
\def\epr{{e-print astro-ph/}}

\begin{opening}

\title{ASTROPHYSICS AT THE HIGHEST ENERGY FRONTIERS}

\author{F.W. STECKER}

\institute{Laboratory for High Energy Astrophysics\\
NASA Goddard Space Flight Center, Greenbelt, MD, USA}

\end{opening}

\begin{document}

\begin{abstract}

I discuss recent advances being made in the physics and astrophysics of
cosmic rays and cosmic \grays\ at the highest observed energies as well as the
related physics and astrophysics of very high energy cosmic neutrinos. I also
discuss the connections between these topics.

\end{abstract}

\section{Introduction}

In these lectures, I will discuss physics and astrophysics at the highest 
energies using current astrophysical observations. By taking a synoptic
view of ultrahigh energy hadrons, photons and neutrinos, one can gain
insights into the profound connections between different fields of 
observational astronomy and astrophysics which use different experimental
techniques. Observations have been made of cosmic \grays\ up to 50 TeV
energy and of ultrahigh energy cosmic rays up to 300 EeV ($3 \times 10^8$ TeV).
As of this date, no very high or ultrahigh energy cosmic neutrinos have been
detected, however, the {\it AMANDA} (Antarctic Muon and Neutrino Detector
Array) experiment, now in operation, is searching for neutrinos above 1 TeV
energy (Wischnewski 2002).

The subjects of these lectures concern some of the deepest questions in
frontiers of cutting-edge astrophysics. They also involve physics at the
highest energy frontiers. The new physics which has been and may be invoked
to explain the highest energy observations comprise such questions as the
violation of Lorentz invariance (special relativity), grand unification of
the electroweak and strong interactions, quantum gravity theory and the
question of whether we live in a universe containing new large extra
dimensions.
    
\section{The Highest Energy Gamma Rays}

The highest energy \grays\ observed to date were produced by the Vela pulsar 
and by PSR 1706-44, by supernova remnants in our galaxy 
and by extragalactic sources known as blazars. Blazars are a class active 
galaxies believed to have supermassive black holes in their nuclei which 
gravitationally power jets which produce
massive amounts of nonthermal radiation. They are distinguished by the 
condition that their jets are pointed almost directly at us, producing
interesting relativistic effects such as rapid time variability and Lorentz
boosted radiation and also superluminal motion in many cases 
(Jorstad \etal 2001).

\subsection{Pulsars}

There are two theoretical models which have been proposed to account for the 
origin of pulsed \gray\ emission in pulsars, {\it viz.}, ``the outer gap 
model'', where particle acceleration occurs in the outer magnetosphere of the 
pulsar (Cheng, Ho and Ruderman 1986), 
and ``the polar cap model'', where particle acceleration occurs near the 
magnetic polar cap of the pulsar (Daugherty and Harding 1996). 

In the outer gap model, in the case of the Crab pulsar, the resulting 
electron-positron pairs can generate TeV \grays\ through the 
synchrotron-self-Compton (SSC) mechanism. In SSC emission models, the source 
has a natural two-peaked spectral energy distribution. The lower energy peak 
is produced by synchrotron radiation of relativistic electrons accelerated in 
the source and the higher energy peak is produced by these same electrons 
upscattering the synchrotron component photons to TeV energies by Compton 
interactions. In the case of other pulsars, the low energy photons may come
from curvature radiation of electrons and positrons as they follow the 
magentic field lines of the pulsar. This curvature radiation would then be
Compton upscattered. 

In the polar cap model, the strong magnetic field near the pulsar would 
cut off any TeV photons emitted near the surface of the pulsar {\it via} 
single-photon electron-positron pair production off the magnetic field 
(Erber 1966). 
Thus, the detection of TeV pulsed emission from 
ordinary pulsars would favor the outer gap model.
However, for millisecond pulsars, whose {\it B}-fields are 3-4 orders of
magnitude lower than is the case for regular pulsars, the high energy
cutoff from single-photon pair production is 3-4 orders of magnitude higher 
in energy (Bulik \etal 2000).
Thus, TeV emission may be detected in these 
sources in the future even in the case of the polar cap \gray\ production.

\subsection{Supernova Remnants}

Very high energy (TeV) \grays\ have been reported from several supernova 
remnants, {\it viz.}, the Crab Nebula (Weekes \etal 1989), 
the Vela pulsar wind nebula (Yoshikoshi\ \etal 1997), 
and the nebulae associated with PSR 1706-44 (Kifune\ \etal 1995), 
Cassiopeia A (P\"{u}hlhofer\ \etal 1999), 
SN1006 (Tanimori \etal 1989), 
and RXJ1713.7-3946 (Muraishi\ \etal 2000). 

The SSC mechanism has been invoked to account for the TeV emission in the
Crab Nebula (de Jager and Harding 1992).
Another mechanism which has been proposed to explain the TeV emission in 
supernova remnants is the Compton
scattering of very high energy relativistic electrons off the 2.7 K cosmic
background radiation. This has been proposed for specifically for the remnant
SN1006 (\eg , Pohl \etal\ 1996; Mastichiadas and de Jager 1996).

There is also the important possibility that TeV \grays\ could be produced
in supernova remnants by accelerated highly relativistic {\it protons} 
interacting with interstellar gas nuclei in the vicinity of the remnant
and producing very high energy \grays\ through the mechanism of the
production and decay of neutral pions ($\pi^0$'s)
(Drury, Aharonian and V\"{o}lk 1994; Gaisser, Protheroe and Stanev 1998).
Almost 70 years ago, Baade and Zwicky (1934) 
first proposed that supernovae
could provide the energy for accelerating cosmic rays in our galaxy. This 
hypothesis gained observational support two decades later when Shklovskii 
(1953) proposed that relativistic electrons radiating in the magnetic field 
of the Crab Nebula produced its optical continuum radiation. Indirect support
for proton acceleration in supernovae came from \gray\ observations in the 
1970s (Stecker 1975; 1976). 
It now appears that evidence for the hadronic $\pi^0$ production mechanism
may have been found for the source RXJ1713.7-3946 (Enomoto\ \etal 2002).
While the evidence is not definitive (Butt\ \etal 2002), we can hope for
a resolution with future observations. The detection of high energy neutrinos
from RXJ1713.7-3946 would serve as a smoking gun for hadronic producion since
these neutrinos would be produced by the decay of $\pi^{\pm}$ mesons 
(Alvarez-M\~{u}niz and Halzen 2002).

\subsection{Blazars}

Blazars
were first discovered as the dominant class of extragalactic high energy
\gray\ sources by the {\it EGRET} detector on the Compton Gamma Ray Observatory
({\it CGRO}). Because they are at large extragalactic distances, their
spectra are predicted to be modified by strongly redshift dependent absorption 
effects caused by interactions of these $\gamma$-rays with photons of the 
intergalactic IR-UV background radiation (Stecker, de Jager and Salamon 1992).

The highest energy extragalactic \gray\ sources 
are those blazars  known as X-ray selected BL Lac objects (XBLs), or
alternatively as high frequency BL Lac objects (HBLs). They are expected to 
emit photons in the multi-TeV energy range 
(Stecker, de Jager and Salamon 1996), but  only the nearest ones
are expected to be observable, the others being hidden by intergalactic
absorption (Stecker \etal 1992).

There are now $\sim$70  ``grazars'' ($\gamma$-ray blazars) which have been 
detected by the {\it EGRET} team at GeV energies (Hartman\ \etal\ 1999)
These sources, optically violent variable quasars
and BL Lac objects, have been detected out to a redshift of $\sim$2.3.

In addition, several TeV grazars have been discovered at low redshifts
($z < 0.13$), {\it viz.}, Mkn 421 at $z = 0.031$ (Punch\ \etal 1992),
Mkn 501 at $z = 0.034$ (Quinn\ \etal (1996),
1ES2344+514 at $z = 0.44$ (Catanese\ \etal 1998), 
PKS 2155-304 at $z = 0.117$ (Chadwick\ \etal 1998),
and 1ES1426+428 (a.k.a. H1426+428) at $z = 0.129$ 
(Aharonian\ \etal 2002; Horan\ \etal 2002). These sources all fit the
two candidate source criteria of Stecker\ \etal (1996) of closeness and a 
high frequency synchrotron peak which is prominant in X-ray emission. 
Recently, another criterion has been suggested for TeV candidate sources, 
{\it viz.} a relatively large flux in the synchrotron peak (Costamante
and Ghisellini 2001). The closeness criterion has to do with
extragalactic TeV \gray\ absorption by pair production (Stecker\ \etal 1992).
The other two criteria have to do with the synchrotron-self-Compton (SSC)
mechanism believed to be primarily responsible for the TeV emission of the
HBL sources. In the SSC models, an HBL blazar has a 
two-peaked spectral energy distribution (see pulsar section above)
with the lower energy peak in the radio to X-ray range and the higher 
energy peak in the X-ray to multi-TeV \gray\ range.

Very high energy \gray\ beams from blazars can be used to 
measure the intergalactic infrared radiation field, since 
pair-production interactions of \grays\ with intergalactic IR photons 
will attenuate the high-energy ends of blazar spectra (Stecker \etal 1992). 
In recent years, this concept has been used successfully to place upper limits 
on the the diffuse infrared radiation background (DIRB) (Stecker and de Jager
1993, 1997; Dwek 1994; Stanev and Franceschini 1997; Biller \etal 1998).
Determining the DIRB, in turn, allows us to 
model the evolution of the galaxies which produce it. 
As energy thresholds are lowered in both existing and planned ground-based
air \v Cerenkov light detectors and with the launch of the {\it GLAST}
(Gamma-Ray Large Area Space Telescope) in 2006, cutoffs in the \gray\ 
spectra of 
more distant blazars are expected, owing to extinction by the DIRB. These
can be used to explore the redshift dependence of the 
DIRB (Salamon and Stecker 1998 (SS98)).

\subsection{The Diffuse Low Energy Photon Background and
Extragalactic Gamma Ray Absorption}

The formulae relevant to absorption calculations involving pair-production 
interactions with redshift factors are given in Stecker\ \etal (1992).
For $\gamma$-rays in the TeV energy range, the pair-production cross section 
is maximized when the soft photon energy is in the infrared range:
\begin{equation} 
\lambda (E_{\gamma}) \simeq \lambda_{e}{E_{\gamma}\over{2m_{e}c^{2}}} =
1.24E_{\gamma,TeV} \; \; \mu m 
\end{equation}
where $\lambda_{e} = h/(m_{e}c)$ 
is the Compton wavelength of the electron.
For a 1 TeV $\gamma$-ray, this corresponds to a soft photon having a
wavelength $\sim$ 1 \mic. (Pair-production interactions actually
take place with photons over a range of wavelengths around the optimal value as
determined by the energy dependence of the cross section.) 
If the emission spectrum of an extragalactic source extends beyond 20 TeV, 
then the extragalactic infrared field should cut off the {\it observed} 
spectrum between $\sim 20$ GeV and $\sim 20$ TeV, depending on the redshift 
of the source (Stecker and Salamon 1997; SS98).

\begin{figure}
\centerline{\psfig{file=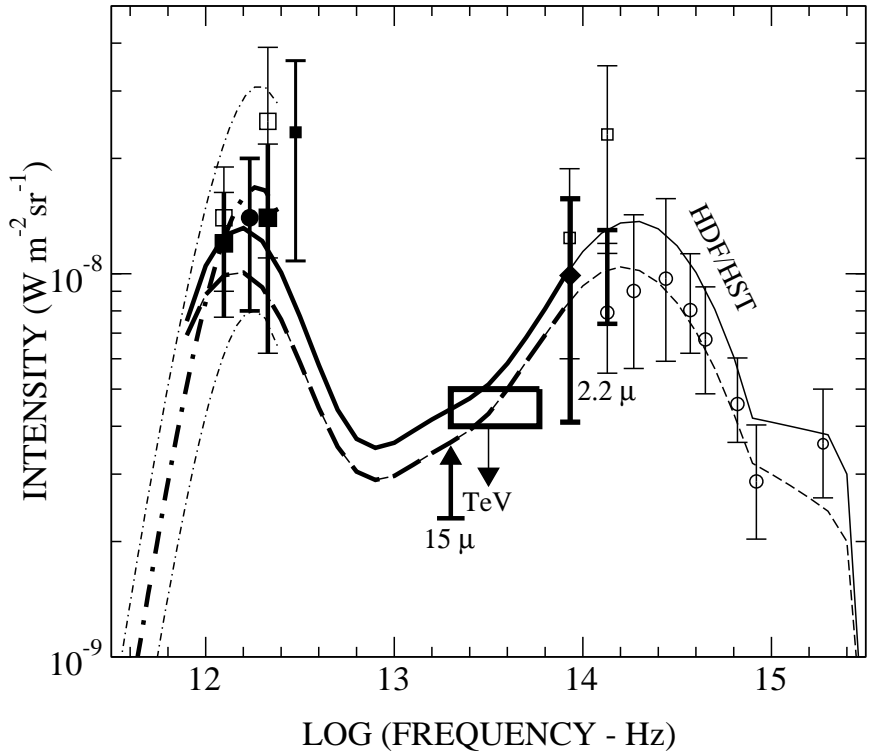,width=13.0truecm}}
\vspace{-8cm}
\caption{The SED of the EBL (see text for references and descriptions). 
All error bars given at the $\pm 2\sigma$ level. 
The Malkan and Stecker (2001) fast evolution model is shown by the upper, 
thick solid curve
between $\log_{10}\nu=11.8$ to 13.8) and their baseline model is shown by
the thick dashed line over the same frequency range.
Convergent Hubble Deep Field galaxy counts (Madau and Pozzetti 2001): 
open circles; Ground--based galaxy counts limits at $2.2\mu$m 
(see text): thick vertical bar marked ``$2.2\mu$'';
{\it COBE-DIRBE} photometric sky residuals (Wright and Reese 2000): small 
open squares;
TeV $\gamma$-ray-based upper limits: thick box marked ``TeV'' (see text for
references);
{\it ISOCAM} lower limit at $15\mu$m: upward arrow marked $15\mu$ (Elbaz
{\it et al.} 1999);
{\it COBE-DIRBE} far-IR sky residuals (Hauser {\it et al.} 1998): large 
open squares at 140 $\mu$m and 240 $\mu$m;
{\it COBE-DIRBE} data recalibrated using the {\it COBE- FIRAS} calibration 
(see text): large solid
squares without error bars; {\it ISOCAM} 170 $\mu$m flux (Kiss {\it et al.}
2001): solid circle; 
{\it COBE-FIRAS} far-IR sky residuals (Fixsen, {\it et al.} 1998): 
thick dot-dash curve 
with $\sim 95\%$ confidence band (thin dot--dash band);
$100 \mu$m {\it COBE-DIRBE} point (Lagache {\it et al.} 2000): 
small solid square; flux at 3.5 $\mu$m from {\it COBE-DIRBE} 
(Dwek and Arendt 1998): solid diamond.}
\label{abs501f1}
\end{figure}

Several attempts have been made to infer the IR SED (spectral energy 
distribution) of the DIRB either from model
calculations or observations. (See Hauser and Dwek (2001) for the latest 
review.) Such information can be used to calculate the optical 
depth for TeV range photons as a function of energy and redshift.
Figure \ref{abs501f1} 
summarizes the observationally derived values for the extragalactic
optical-UV, near-IR and far-IR fluxes which now exist. We will refer to
the totality of these fluxes as the extragalactic background light (EBL) 
Unfortunately, foreground emission prevents the direct detection of the 
EBL in the mid-IR wavelength range. (See discussion in Hauser and
Dwek 2001.) However, other theoretical models such as those of 
Tan, Silk and Balland (1999), Rowan-Robinson (2000) and Xu (2000) predict 
fairly flat SEDs in the mid-infrared range with average flux levels in the 
3 to 4 nW m$^{-2}$sr $^{-1}$ as do the Malkan and Stecker (2001) models shown 
in Figure \ref{abs501f1}. These flux levels are also 
sconsistent with the indirect mid-IR 
constraints indicated by the box in Figure \ref{abs501f1}. 
(These constraints are summarized by Stecker (2001).)
\footnote{We note that the {\it COBE-DIRBE} group has argued that a real 
flux derived from the {\it COBE-DIRBE} data at 100 $\mu$m, as 
claimed by Lagache {\it et al.} (2000), is untenable because isotropy in the 
residuals (after foreground subtractions)
could not be proven. Dwek {\it et al.} (1998) have concluded that only a
conservative lower limit of 5 m$^{-2}$sr$^{-1}$ could be inferred at
100 $\mu$m.} 

\hskip 5cm

The two Malkan and Stecker (2001) SED curves for the DIRB,
extended into the optical-UV range by the hybrid model of de Jager and 
Stecker (2002) (DS02), give a reasonable 
representation of the EBL in the UV to far-IR. Other EBL models in the 
literature whose flux levels roughly fit the present data and have the same 
spectral characteristics, {\it i.e.}, a stellar optical peak, a far-IR dust 
emission peak, and a mid-IR valley which allows for some warm dust emission
(see review of Hauser and Dwek 2001), 
should give similar results on the optical depth of the near Universe 
$\tau(E,z)$ to high energy $\gamma$-rays. 

\begin{figure}
\centerline{\psfig{file=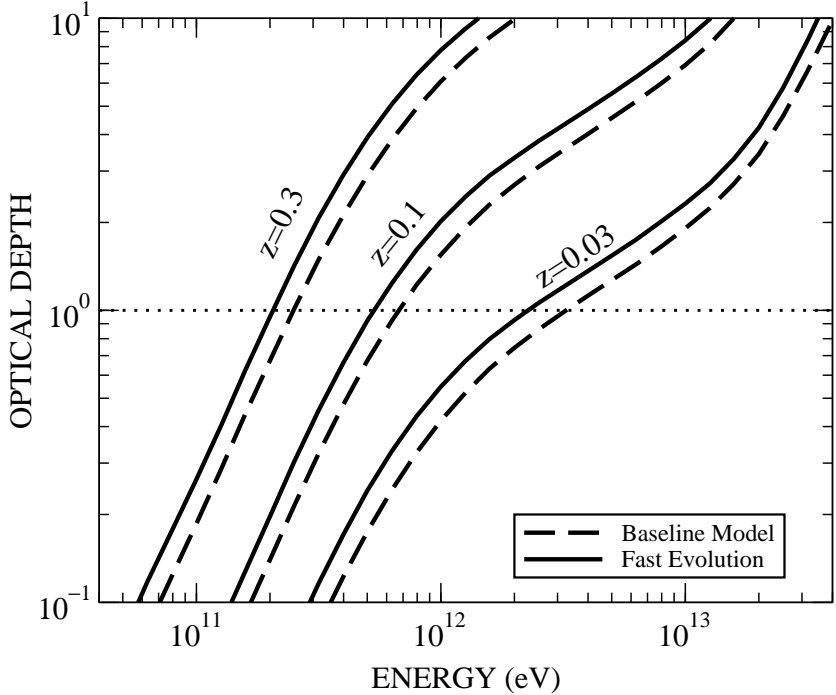,width=13.0truecm}}
\vspace{-8cm}
\caption{The optical depth for $\gamma$-rays above 50 GeV
given for redshifts between 0.03 and 0.3 (as labelled) calculated using the 
medium (dashed lines) and fast evolution (solid lines)
SED of Malkan and Stecker (2001).} 
\label{abs501f2}
\end{figure}

DS02 rederived the optical depth of the universe 
to high energy $\gamma$-rays as a function of energy and 
redshift for energies betweeen 50 GeV and 100 TeV and redshifts between 0.03 
and 0.3 using their derived hybrid EBL SEDs.
Figure \ref{abs501f2} shows $\tau(E,z)$ calculated using the 
baseline and fast evolution hybrid models of DS02 for $\gamma$-ray energies
down to $\sim 50$ GeV, which is the approximate threshold energy for 
meaningful image analyses in next generation ground based 
$\gamma$-ray telescopes such as {\it MAGIC, H.E.S.S.} and {\it VERITAS}. 
Where they overlap, the DS02 results are in good 
agreement with the metallicity corrected results of SS98,which give the 
optical depth of the universe to \grays\ out to a redshift of 3 and extend 
to lower energies (See Section 2.7.)

\hskip 5cm

DS02 also obtained parametric expressions for $\tau(E,z)$ 
for $z<0.3$ and $0.1<E_{\rm TeV}<50$ which are accurate to within 5\% and
can be used for numerical calculations. 

\subsection{The Nearby TeV Blazar Source Mkn 501}

The nearby blazar Mkn 501 is of particular interest because a very
detailed determination of its spectrum was obtained by observing it while
it was strongly flaring in 1997. The spectrum observed at that time by the 
{\it HEGRA} air \v Cerenkov telescope system (Aharonian, {\it et al.}  
2001a) extended to energies greater than 20 TeV, the highest energies yet 
observed from an extragalactic source.

Using this observational data, DS02 derived the {\it intrinsic} $\gamma$-ray 
spectrum of Mkn 501 during its 1997 flaring state by compensating for the
effect of intergalactic absorption. They found that the time averaged
spectral energy distribution of Mkn 501 while flaring had a broad, 
flat peak in the $\sim$5--10 TeV range which corresponds
to the broad, flat time averaged X-ray peak in the $\sim$ 50--100 keV range 
observed during the flare.
The spectral index of the derived intrinsic differential photon spectrum for
Mkn 501 at energies below $\sim$2 TeV was found to be 
$\sim$1.6--1.7. This corresponds to a time averaged spectral
index of 1.76 found in soft X-rays at energies below the X-ray (synchrotron) 
peak (Petry {\it et al.} 2000). These results appear to favor an
SSC origin for the TeV emission together with jet 
parameters which are consistent with time variability constraints 
within the context of a simple SSC model. 
The similarity of the soft X-ray and $\sim$TeV spectral indices of the
two components of the source spectrum implies that $\gamma$-rays below 
$\sim 1$ TeV are produced in the Thomson regime by scattering off synchrotron 
photons with energies in the optical-IR range. On the other hand, 
$\gamma$-rays near the $\sim$ 7 $\pm$ 2 TeV Compton peak appear to be the 
result of scattering in the Klein-Nishina range.

\hskip 5cm
\begin{figure}
\centerline{\psfig{file=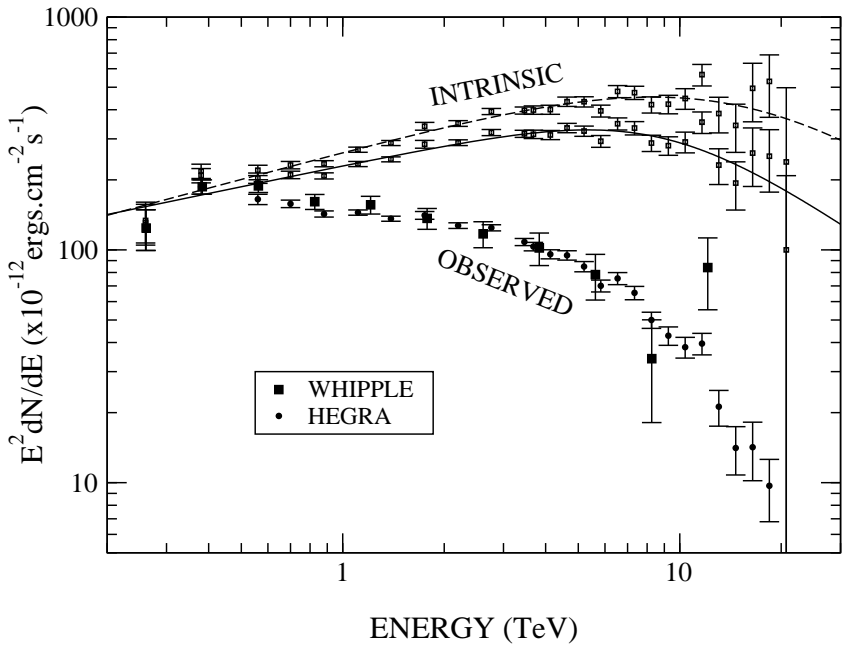,width=13.0truecm}}
\vspace{-8cm}
\caption{The observed spectrum and derived intrinsic spectrum
of Mkn 501. The observed spectral data are as measured by {\it HEGRA} 
(solid circles) and {\it Whipple} (solid squares). The upper points
are the absorption corrected data points (marked ``INTRINSIC'') using our
fast evolution hybrid EBL (upper data set and solid curve fit) and baseline
hybrid EBL (lower data set with dashed curve fit).} 
\label{abs501f3}
\end{figure}

The observed spectrum between 0.56 and 21 TeV was 
obtained from contemporaneous observations of Mkn 501 by the 
{\it HEGRA} group (Aharonian {\it et al.} 2001a) and the {\it Whipple} group 
(Krennrich {\it et al.} 1999).
These observations are consistent with each other (within errors)
in the overlapping energy range between 0.5 and 10 TeV, 
resulting in a single spectrum extending
over two decades of energy, marked ``OBSERVED'' in Figure \ref{abs501f3}.

Using both the {\it Whipple} and {\it HEGRA} data and correcting for 
absorption by multiplying by $e^{\tau(E)}$ evaluated at $z = 0.034$
with their newly derived values for the opacity,
DS02 derived the intrinsic spectrum of Mkn 501 over two decades of energy.
This is given by the data points and two curves marked ``INTRINSIC'' in
Figure \ref{abs501f3}. The upper of these curves corresponds to the fast 
evolution case; the lower curve corresponds to the baseline model case.  

Fossati {\it et al.} (2000) has suggested a parameterization to describe 
smoothly curving blazar spectra. This parameterization is of the form

\begin{equation}
\frac{dN}{dE}=
KE^{-\Gamma_1}\left(1+(\frac{E}{E_B})^f\right)^{(\Gamma_1-\Gamma_2)/f}
\end{equation}
A  spectrum of this form
changes gradually from a spectral index of $\Gamma_1$
to an index of $\Gamma_2$ when the energy $E$ increases through the
break energy $E_B$. The parameter $f$ describes the rapidity (``fastness'')
of the change in spectral index over energy. 
DS02 applied the formalism of Fossati {\it et al.} 
(2000) to their Mkn 501 intrinsic flare spectrum and found best-fits for the 
parameters $K$, $E_B$, $f$, $\Gamma_1$ and $\Gamma_2$ after correcting the
observed spectrum for intergalactic absorption. 

Whereas the low energy spectral index
$\Gamma_1$ was found to be well constrained, the
higher energy index $\Gamma_2$ is unconstrained.
The SED peaks at $E_{M} \sim 8-9$ TeV (independent of the unconstrained 
$\Gamma_2$) in the case where the fast evolution EBL is assumed and that
$E_{M} \sim 5$ TeV if the baseline EBL is assumed. (See Figure 
\ref{abs501f3})

Since Mkn 501 is a giant elliptical 
galaxy with little dust, it is reasonable to assume that the galaxy itself
does not produce enough infrared radiation to provide a significant opacity
to high energy $\gamma$-rays. Such BL Lac objects have little gas (and 
therefore most likely little dust) in their nuclear regions. It also appears
that \gray\ emission in blazars takes place at superluminal knots in 
the jet downstream of the core and at any putative accretion disk.
So, if the EBL SED of DS02 is approximately
correct, it is reasonable to assume that the dominant absorption process is
intergalactic and that pair-production in the jet in negligible. This
hypothesis is supported by the fact that the high energy $\gamma$-ray SED
did not steepen during the flare.
This implies that the optical depth given by eq. (4)
is less than unity out to the highest observed energy $E \sim
20$ TeV. Thus, it appears that the high energy turnover in the observed
Mkn 501 \gray\ spectrum above $\sim$ 10 
TeV can be understood solely as a result of intergalactic absorption. 

The intrinsic SED of Mkn 501 derived by DS02 is quite 
flat in the multi-TeV range as shown in Figure 3. This is in 
marked contrast to the dramatic turnover in its 
observed SED. This is strong evidence that the
observed spectrum shows just the absorption effect predicted. We will see
that the spectral observations of other blazars, although not nearly as
good in most cases, also exhibit evidence of intergalactic absorption.

\subsection{Absorption in the Spectra of other Blazars}

Observations of Mkn 421, the closest TeV blazar which has a redshift very
similar to that of Mkn 501, were made by the {\it Whipple} group up to an
energy of 17 TeV (Krennrich \etal 2002). Their results indicate a turnover in 
the spectrum 
at energies above $\sim$4 TeV caused by extragalactic absorption, quite
similar to that observed in the spectrum of Mkn 501 (see above).

\begin{figure}[t]

\centerline{\psfig{file=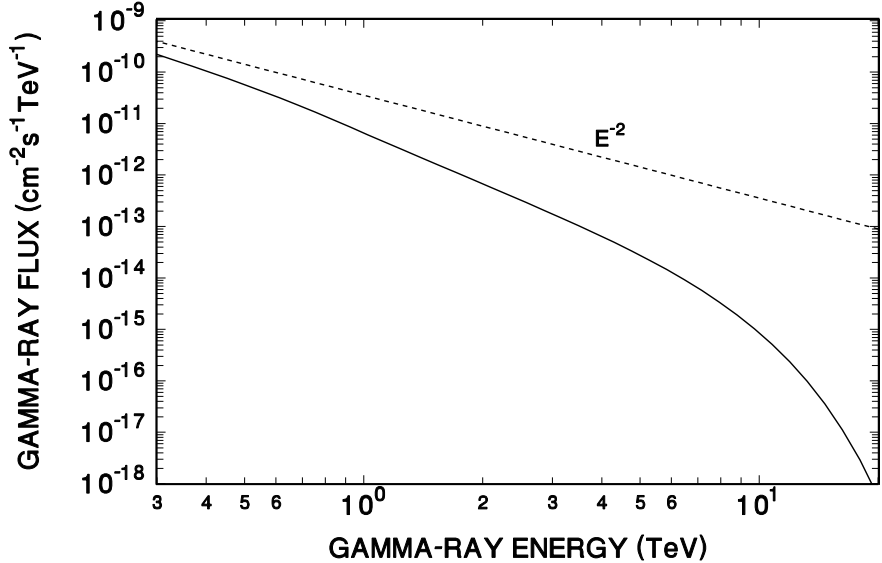,width=13.0truecm,angle=270}}
\vspace{2.0truecm}
\caption{Predicted differential absorbed spectrum, for PKS 2155-304 
(solid line) assuming an $E^{-2}$ differential source spectrum (dashed line) 
normalized
to the integral flux given by Chadwick \etal (1998).}
\label{2155}
\end{figure}

As to sources at somewhat higher redshifts,
Stecker (1999) considered the blazar source PKS 2155-304, located at a moderate
redshift of 0.117, which has been reported by the Durham group to have
a flux above 0.3 TeV of $\sim 4 \times 10^{-11}$ cm$^{-2}$ s$^{-1}$,
close to that predicted by a simple SSC model.
Using absorption results obtained by Stecker and de Jager (1998)
and assuming an $E^{-2}$ source spectrum, Stecker (1999) predicted an 
absorbed (observed) spectrum as shown in Figure \ref{2155}. As indicated in 
the figure, it was predicted that this source should have its spectrum 
steepened by $\sim$ 1 in its spectral index between $\sim 0.3$ and $\sim 3$ 
TeV and should show a pronounced absorption turnover above $\sim 6$ TeV.

There are now recent observations of the spectrum of another TeV blazar
1ES1426+428 at a redshift of 0.129, not too different from that of 
PKS 2155-304. Interestingly, it has been found by three groups that the 
spectrum of this source is quite steep, with a photon differential spectral 
index greater than 3 (Aharonian\ \etal 2002; Petry\ \etal 2002; Djannati-Atai,
\etal 2002). This steep spectrum, similar to that predicted by Stecker 
(1999) for PKS2155-304 at a redshift of 0.117, is more evidence for 
extragalactic absorption.

\subsection{The Gamma-Ray Opacity at High Redshifts}

Salamon and Stecker (1998) (SS98) have calculated 
the \gray\ opacity as a function of both energy and redshift
for redshifts as high as 3 by taking account of the evolution of both the
SED and emissivity of galaxies with redshift. 
In order to accomplish this, 
they adopted the recent analysis of Fall\ \etal\ (1996) and 
also included the effects of metallicity evolution on galactic SEDs.
They then gave predicted \gray\ spectra 
for selected blazars and extend our calculations of the 
extragalactic \gray\ background from blazars to an energy of 500 GeV with
absorption effects included. 
Their results indicate that the extragalactic
\gray\ background spectrum from blazars should steepen significantly 
above 15-20 GeV, owing to extragalactic absorption. Future observations of
such a steepening would thus provide a test of the blazar origin
hypothesis for the \gray\ background radiation. 

SS98 calculated stellar
emissivity as a function of redshift at 0.28 $\mu$m, 0.44 $\mu$m,
and 1.00 $\mu$m, both with and without a metallicity correction.
Their results agree well with the emissivity obtained by the Canada-France
Redshift Survey (Lilly \etal\ 1996)
over the redshift range of the 
observations ($z \le 1$).
The stellar emissivity in the universe is found to peak
at $ 1 \le z \le 2$, dropping off steeply at lower reshifts and is roughly
constant higher redshifts (\eg Steidel 1999). Indeed,
Madau and Schull (1996) have used observational data from the 
Hubble Deep Field to show that metal production has a similar redshift 
distribution, such production being a direct measure of the star formation 
rate. (See also Pettini \etal 1994.)

\begin{figure}
\centerline{\psfig{file=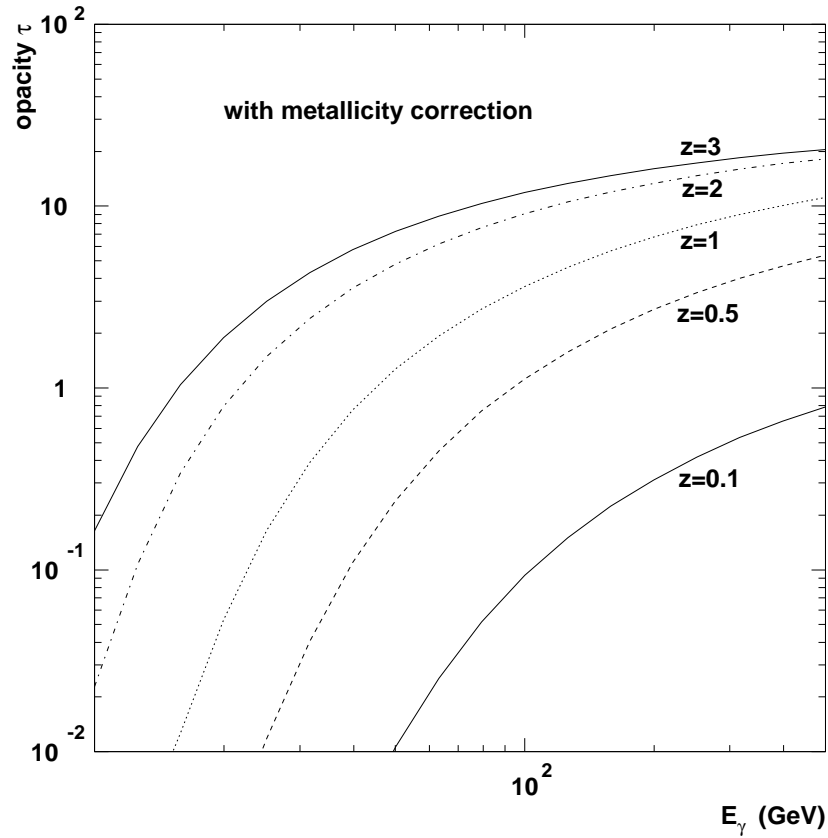,width=13.0truecm}}
\vspace{-3cm}
\caption{The optical depth for $\gamma$-rays as a function of energy for
various redshifts calculated using the 
metallicity correction of SS98.}
\label{highzopac}
\end{figure}

The optical depth of the universe to \grays\ as a function of energy
at various redshifts out to $z = 3$ which was derived by SS98 is shown in 
Figure \ref{highzopac}.

\subsection{The Effect of Absorption on the Spectra of Blazars and the
Gamma-Ray Background}

\begin{figure}
\centerline{\psfig{file=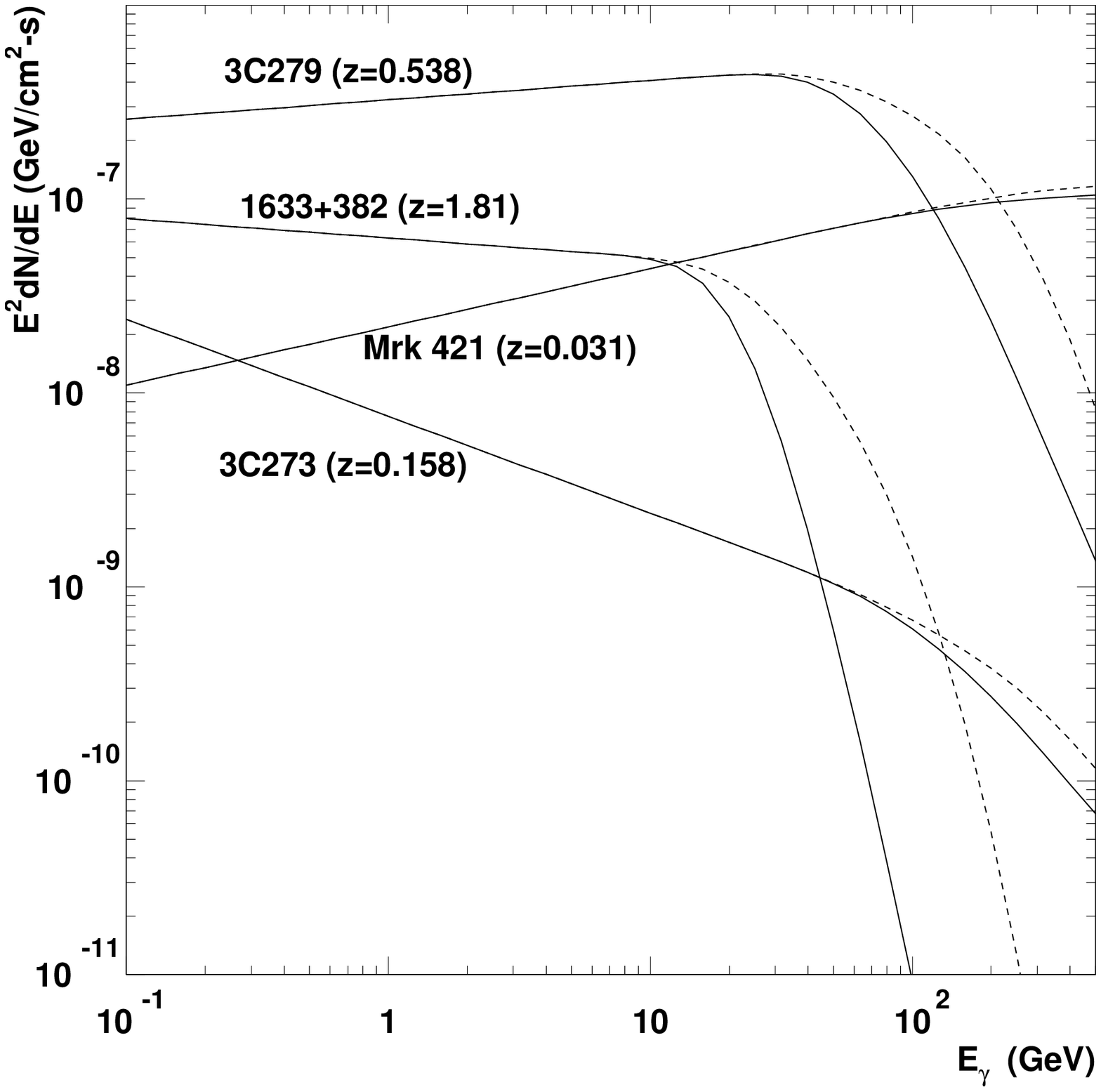,width=12.0truecm}}
\vspace{-3cm}
\caption{ The effect of intergalactic absorption by 
pair-production on the power-law spectra of
four prominent grazars: 1633+382 ($z=1.81$), 
3C279 ($z=0.54$), 3C273 ($z=0.15$), and Mkn 421 ($z=0.031$)
assuming an extrapolation of the spectral indeces for these 
sources measured by {\it EGRET} to high energy.} 
\label{blazarabs}
\end{figure}

With the \gray\ opacity $\tau(E_{0},z)$ calculated out to
$z = 3$ (see previous section), the cutoffs in blazar \gray\ spectra caused 
by extragalactic pair 
production interactions with stellar photons can be predicted.
The left graph in Figure \ref{blazarabs} from SS98
shows the effect of the intergalactic radiation
background on a few of the grazars 
observed by {\it EGRET}, {\it viz.}, 1633+382, 3C279, 3C273, and Mkn 421,
assuming that the mean spectral indices obtained for these sources by
{\it EGRET} extrapolate out to higher energies attenuated only by 
intergalactic absorption.  In considering this figure, it should be noted
that observed cutoffs in many grazar spectra 
may also be affected by natural cutoffs in their source spectra 
(Stecker, de Jager and Salamon 1996) and intrinsic absorption 
may also be important in some sources (Protheroe and Biermann 1996).

~

\begin{figure}
\centerline{\psfig{file=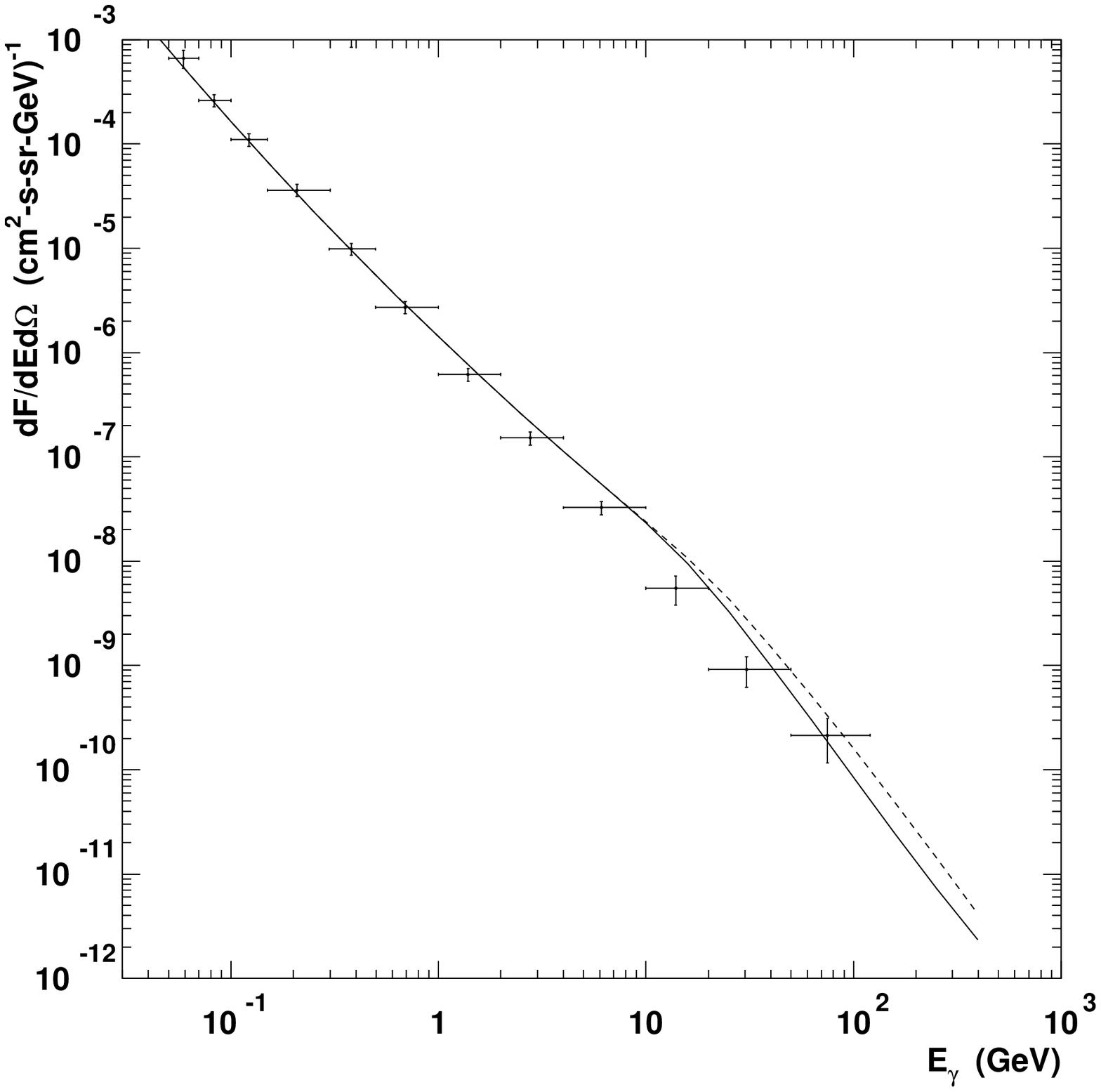,width=12.0truecm}}
\vspace{-3cm}
\caption{ The extragalactic \gray\ background spectrum predicted by the 
unresolved blazar model of Stecker and Salamon (1996)  
with absorption included, calculated for a mean {\it EGRET} point-source 
sensitivity of $10^{-7}$ cm$^{-2}$s$^{-1}$, compared with the {\it EGRET}
data on the \gray\ background (Sreekumar \etal 1998).  
The solid (dashed) curves are calculated with (without) the metallicity
correction function (from SS98).}
\label{gammabackabs}
\end{figure}

Figure \ref{gammabackabs}
shows the background spectrum predicted from unresolved blazars
(Stecker and Salamon 1996a; SS98),
compared with the {\it EGRET} data (Sreekumar\ \etal\ 1998). 
Note that the predicted spectrum steepens 
above 20 GeV, owing to extragalactic absorption by pair-production 
interactions with radiation from external galaxies, particularly at
high redshifts.

Again we note that the predicted background above 10 GeV from
unresolved blazars is uncertain because many blazars are expected to have 
intrinsic cutoffs in their \gray\ production spectra
and by intrinsic \gray\ absorption within such sources is also a possibility. 
Thus, above 10 GeV the 
calculated background flux from unresolved blazars shown in 
Figure \ref{gammabackabs} may actually be an upper limit.
Whether cutoffs in grazar spectra are 
primarily caused by intergalactic absorption can be determined by
observing whether the grazar cutoff energies have the 
type of redshift dependence predicted here.

\subsection{Constaints on the Redshifts of Gamma-Ray Bursts}

The results of the SS98 absorption calculations can also be used to place 
limits on the redshifts (or distances) of \gray\ bursts. 
On 17 February 1994, the {\it EGRET} telescope observed a \gray\ burst
which contained a photon of energy $\sim$ 20 GeV (Hurley\ \etal 1994).
If one adopts the opacity results which include the
metallicity correction, the highest energy photon in this burst 
would be constrained probably to have originated at a redshift
less than $\sim$2. 

On 17 April 1997, the ground based ``Milagrito detector''
observed the burst GRB970417a with an effective threshold of energy of
0.65 TeV (Atkins\ \etal\ 2002).
Using the opacity curves derived by DS02 as shown in Figure 
\ref{abs501f2}, 
we can obtain constraints on the maximum redshift of this GRB. Since the 
observed burst contained photons of energies down to 0.65 TeV, the maximum 
redshift of GRB970417a was $z_{max} \simeq 0.1$.

Future detectors such as {\it GLAST} (Bloom 1996)
should be able to place redshift constraints on bursts observed at higher 
energies. Such constraints may further help to identify the host galaxies 
of \gray\ bursts.

\subsection{Constraints on the Violation of Lorentz Invariance.}

Lorentz invariance violation can be described quite simply in terms of 
different maximal attainable velocities of different particle species as
measured in the preferred frame (Coleman and Glashow 1999).  
Following the well-defined formalism for 
LI violation discussed by Coleman and Glashow (1999), (see also
Colladay and Kostelecky 1998), 
the maximum attainable velocity of an
electron need not equal the {\it in vacua\/} velocity of light,
{\it i.e.,}  $c_e \ne c_\gamma$. The physical
consequences of this violation of LI depend on the sign of the difference. 
Defining

\begin{equation}
c_{e} \equiv c_{\gamma}(1 +  \delta) ~ , ~ ~~~0< |\delta| \ll 1\;,  
\end{equation}

\noindent
Stecker and Glashow (2001) consider the two cases of positive and negative 
values of $\delta$ separately. 

{\it Case I:} If $c_e<c_\gamma$ ($\delta < 0$), 
the decay of a photon into an electron-positron pair is kinematically allowed
for photons with energies exceeding

\begin{equation}
E_{\rm max}= m_e\,\sqrt{2/|\delta|}\;. 
\end{equation}

\noindent
The  decay would take place rapidly, so that photons with energies exceeding 
$E_{\rm max}$ could not be observed either in the laboratory or as cosmic 
rays. From the fact that photons have been observed with energies   
$E_{\gamma} \ge$ 50~TeV from the Crab nebula 
(Tanimori\ \etal\ 1998),
it follows from eq.(9) that $E_{\rm max}\ge 50\;$TeV, 
or that -$\delta < 2\times 10^{-16}$.

{\it Case II:}  Here we are concerned with the remaining possibility, where
 $c_e>c_\gamma$ ($\delta > 0$) and electrons become superluminal if their
energies exceed $E_{\rm max}/2$.
Electrons traveling faster than light will emit light  at all frequencies by a
process of `vacuum \v Cerenkov radiation.' This process occurs rapidly, so
that superluminal electron energies
 quickly approach $E_{\rm max}/2$. 
 However, because electrons have been seen in the cosmic radiation 
with energies up to $\sim\,$1~TeV (Nishimora\ \etal\ 1980),
it follows that  $E_{\rm max} \ge 2$~TeV, which leads to 
an upper limit on $\delta$ 
for this case of $1.3 \times 10^{-13}$. This limit
is  three orders of magnitude weaker than the limit obtained for Case I. 
However, if the observed $\sim$TeV \gray\ emission from the Crab Nebula
is produced by very high energy electrons {\it via} the SSC mechanism
(de Jager and Harding 1992), then the implied maximum electron energy is
$\sim 10^4$ TeV. In this case we would get an {\it indirect} upper limit
on $\delta$ of $1.3 \times 10^{-21}$.

Stecker and Glashow (2001) have also shown how stronger bounds on $\delta$ 
can be set using observations of very high energy cosmic ray photons. For 
case I, the discussion is trivial: The mere detection of cosmic $\gamma$-rays
with energies greater that 50~TeV from sources in our galaxy would 
improve the bound on $\delta$. 
For case II, if LI is broken so that $c_e>c_\gamma$, the threshold energy for 
the pair production process $\gamma + \gamma \rightarrow e^+ + e^-$ is altered.The square of the four-momentum becomes

\begin{equation}
2\epsilon E_{\gamma}(1 - \cos \theta) - 2E_{\gamma}^2\delta = 4\gamma^2m_{e}^2 >4 m_{e}^2
\end{equation}

\noindent where $\epsilon$ is the energy of the low energy (infrared) photon and $\theta$
is the angle between the two photons. The second term on the left-hand-side
comes from the fact that $c_{\gamma} =  
\partial E_{\gamma}/\partial p_{\gamma}$.

For head-on collisions ($\cos \theta = -1$) the minimum low energy photon
energy for pair production becomes 

\begin{equation}
\epsilon_{min} = m_{e}^2/E_{\gamma} +  (E_{\gamma}\,\delta)/2
\end{equation}

\noindent It follows that the condition for a significant 
increase in the energy
threshold for pair production is 

\begin{equation}
E_{\gamma} \ge E_{\rm max} \quad\ \  {\rm or\ equivalently}\quad\ \ 
\delta \ge 2m_e^2/E_{\gamma}^2\,.
\end{equation}

Extragalactic photons with the highest energies yet observed originated in a 
powerful flare coming from Mkn
501 (Aharonian\ \etal\ (1999).
We have seen that the Mkn 501 observations indicate that 
its multi-TeV  spectrum is consistent with what one would expect from 
intergalactic absorption (SD02).
Since there is no significant decrease in the optical 
depth of the universe at the distance of Mkn 501 for $E_{\gamma} \le 20$~TeV,
it then follows from eq. (5) that $\delta \le 2(m_{e}/E_{\gamma})^2 = 
1.3 \times 10^{-15}$. This constraint is two orders of 
magnitude stronger than that obtained from the direct cosmic-ray electron 
data but is six orders of magnitude less than the indirect theoretical
limit obtained from the SSC Crab nebula $\gamma$-ray emission argument given
above.

Future detection of galactic $\gamma$-rays with energies
greater than 50 TeV would strengthen the bound on $\delta$ for Case I. For 
Case II, the detection of cosmic 
$\gamma$-rays above $100(1+z_{s})^{-2}$ TeV from a source
at a redshift $z_{s}$,  would be strong evidence for LI breaking with
$\delta \ge 0$. 
This is because the very large density ($\sim\,$400~cm$^{-3}$) of 3K cosmic 
microwave photons would otherwise absorb \grays\ of energy $\ge 100$ TeV 
within a distance of $\sim$ 10 kpc,  with  this critical energy  
reduced by a factor of $\sim (1+z_{s})^2$ for extragalactic sources at 
redshift $z_{s}$ (Stecker 1969).

The constraints on $\delta$ lead to constraints on quantum gravity theory.
Amelino-Camelia\ \etal (1998) and Ellis\ \etal (1998) have proposed that
the back reaction of the vacuum in their quantum gravity scenario would
lower the effective velocity of a photon or electron so that

\begin{equation}
\delta ~~ \simeq \pm E/M_{QG} + {\rm higher~ order~ terms.} 
\end{equation}

As discussed above, Stecker and Glashow (2001) have shown that 
if $c_e<c_\gamma$ ($\delta < 0$), ({\it Case I}), 
$-\delta \le 2 \times 10^{-16}$ for $E = 5 \times 10^4 {\rm GeV}$,
since 50 TeV \grays\ have been detected from the Crab Nebula. 
According to
eq. (8), this implies that 
$M_{QG} \simeq  -E/\delta  \ge  20 M_{Planck}$,
where $M_{Planck} = 1.2 \times 10^{19}$ GeV, is the Planck mass. 
Since $M_{QG}$ is usually identified with $M_{Planck}$ this may
imply an inconsistency in this scenario. 

For the case when $\delta > 0$ ({\it Case II}), 
the analysis of the Mkn 501 spectrum (see previous discussion) gives an upper
limit on $\delta$ of $1.3 \times 10^{-15}$. This gives the lower limit  
$M_{QG} \ge {\cal{O}}(20TeV/\delta) \sim {\cal{O}}(M_{Planck})$,
implying that the quantum gravity scale in this scenario 
cannot be much less than $M_{Planck}$ (Amelino-Camelia 2002).

\section{The Highest Energy Cosmic Rays}

\subsection{The Data}

\begin{figure}
\centerline{\psfig{figure=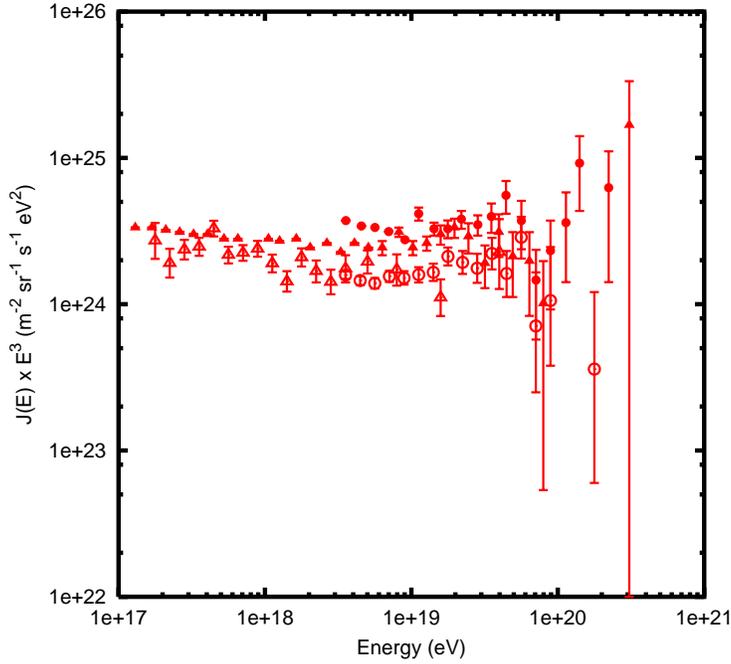,height=14cm}}
\vspace{-1.5cm}
\caption{The ultrahigh energy cosmic ray spectral data from the analysis
of {\it Fly's Eye} (closed triangles), {\it AGASA} (closed circles), 
{\it HiRes} I-monocular (open circles), and {\it HiRes} II-monocular 
(open triangles) observations.}
\label{uhcrdata}

\end{figure}

Figure \ref{uhcrdata} 
shows the data (as of this writing) on the ultrahigh 
energy cosmic ray spectrum from the {\it Fly's Eye, AGASA} and {\it HiRes} 
detectors.\footnote{The AGASA data have been reanalysed and the number of 
events determined to be above 100 EeV has been lowered to eight. (Teshima,
private communication.)} Other data from Havera Park and
Yakutsk may be found in the review by Nagano and Watson (2000) are 
consistent with Figure \ref{uhcrdata}. The new {\it HiRes} data are from
Abu-Zayyad \etal (2002).

For air showers produced by primaries of energies in the 1 to 3 EeV range, 
Hayashida, {\it et al.} (1999) have found a marked directional anisotropy 
with a 4.5$\sigma$ excess from the galactic center region, a 3.9$\sigma$ 
excess from the Cygnus region of the galaxy, and a 4.0$\sigma$
deficit from the galactic anticenter region. This is strong evidence that
EeV cosmic rays are of galactic origin. A galactic plane enhancement in
EeV events was also reported by the Fly's Eye group (Dai\ \etal 1999).

\begin{figure}
\centerline{\psfig{figure=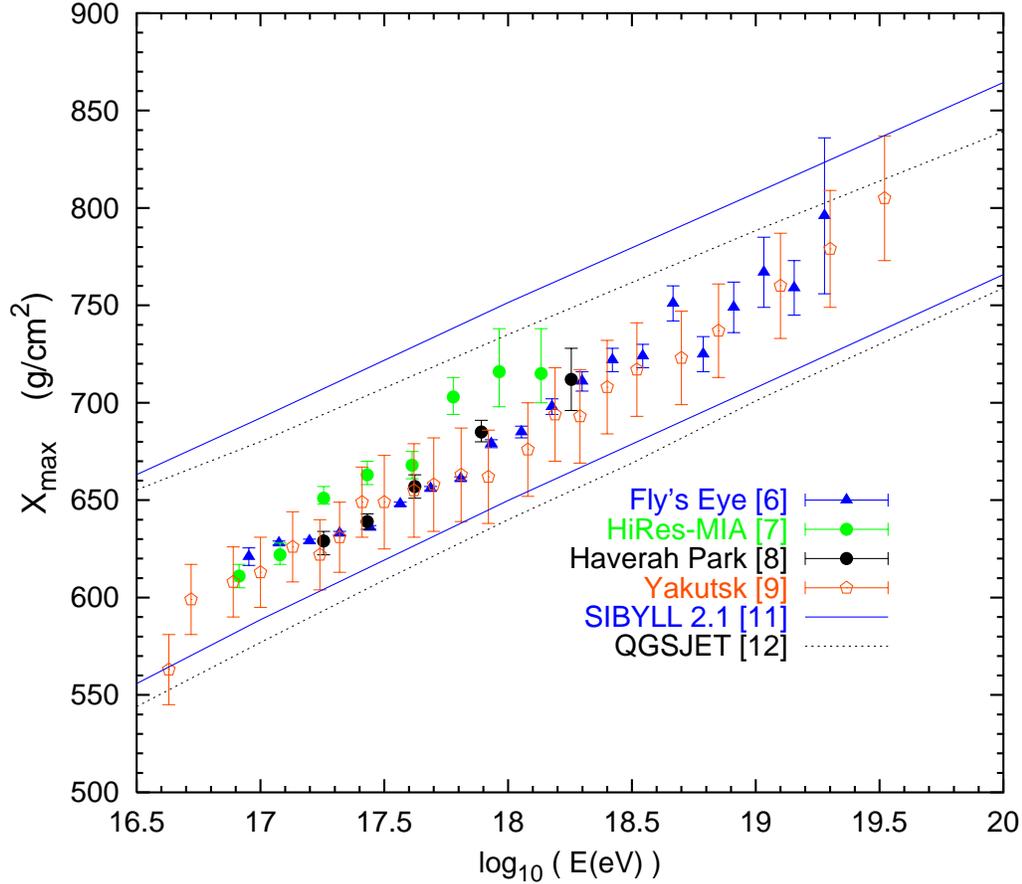,height=12cm}}
\caption{Average depth of shower maximum ($X_{max}$) {\it vs.} energy
compared to the calculated values for protons (upper curves) and Fe
primaries (lower curves) (from Gaisser 2000; 
see references therein).}
\label{parisf2.eps}
\end{figure}

As shown in Figure \ref{parisf2.eps} , at EeV energies, the primary particles
appear to have a mixed or heavy composition, trending toward a light 
composition in the higher energy range around 30 EeV 
(Bird, {\it et al.} 1993; 
Abu-Zayyad, {\it et al.} 2000). This trend, together with 
evidence of a flattening in the cosmic ray spectrum on the 3 to 10 EeV energy 
range 
(Bird, {\it et al.} 1994; 
Takeda {\it et al.} 1998) is evidence for
a new component of cosmic rays dominating above 10 EeV energy. 

The apparent isotropy (no galactic-plane enhancement) of cosmic rays above
10 EeV ({\it e.g.,} Takeda, {\it et al.} 1999), together with
the difficulty of confining protons in the galaxy at 10 to 30 EeV energies,
provide significant reasons to believe that the cosmic-ray component above 
10 EeV is extragalactic in origin. As can be seen from Figure \ref{uhcrdata}, 
this extragalactic component appears to extend to an energy of 300 EeV.
Extention of this spectrum to higher energies is conceivable because such 
cosmic rays, if they exist, would be too rare to have been seen with
present detectors. We will see in the next section that the existence of 300 
EeV cosmic rays gives us a new mystery to solve.

\subsection{The GZK Effect}

Thirty seven years ago, Penzias and Wilson (1965) reported the 
discovery of
the cosmic 2.7K thermal blackbody radiation which was produced very
early on in the history of the universe and which led to the undisputed
acceptance of the ``big bang'' theory of the origin of the universe. Much
more recently, the {\it COBE} (Cosmic Background Explorer) satellite 
confirmed this
discovery, showing that the cosmic background radiation (CBR) has the 
spectrum of the most perfect thermal blackbody known to man. {\it COBE} data 
also 
showed that this radiation (on angular scales $ > 7^\circ$) was isotropic to a 
part in $10^5$ (Mather {\it et al.} 1994). The perfect 
thermal character and smoothness of the CBR proved
conclusively that this radiation is indeed cosmological and that, at the 
present time, it fills the entire universe with a 2.725 K thermal
spectrum of radio to far-infrared photons with a density of 
$\sim 400$ cm$^{-3}$.

\begin{figure}
\centerline{\psfig{figure=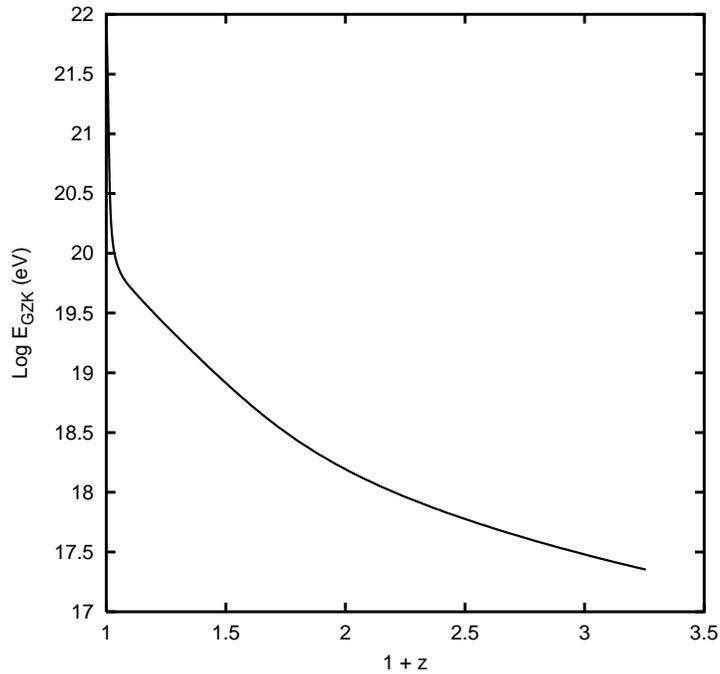,height=14cm}}
\vspace{-1.5cm}
\caption{The GZK cutoff energy versus redshift (Scully 
and Stecker 2002).}
\label{parisf3.eps}
\end{figure}

Shortly after the discovery of the CBR, 
Greisen (1966) and Zatsepin and Kuz'min (1966)
predicted that pion-producing interactions of ultrahigh energy cosmic 
ray protons with CBR photons of target density $\sim$ 400 cm$^{-3}$ should 
produce a cutoff in their spectrum at energies greater than $\sim$ 50 EeV. 
This predicted effect has since become known as the GZK 
(Greisen-Zatsepin-Kuz'min) effect. Following the GZK papers, Stecker (1968)
utilized data on the energy dependence of the photomeson production cross 
sections and inelasticities to calculate the mean energy loss time for protons
propagating through the CBR in intergalactic space as a function of energy.
Based on his results, Stecker (1968) then suggested that the particles of 
energy above the GZK cutoff energy (hereafter referred to as trans-GZK 
particles) must be coming from within the ``Local Supercluster'' of which we 
are a part and which is centered on the Virgo Cluster of galaxies. Thus, the 
``GZK cutoff'' is not a true cutoff, but a supression of the ultrahigh energy
cosmic ray flux owing to a limitation of the propagation distance to a few
tens of Mpc.

The actual position of the GZK cutoff can differ from the 50 EeV predicted
by Greisen (1966). In fact, there could actually be an 
{\it enhancement} at or near 
this energy owing to a ``pileup'' of cosmic rays starting out at higher 
energies and crowding up in energy space at or below the predicted cutoff 
energy (Puget, Stecker and Bredkamp 1976; Hill and 
Schramm 1985; Berezinsky and Grigor'eva 1988; 
Stecker 1989; Stecker and Salamon 1999).
The existence and intensity of this predicted pileup depends critially on the
flatness and extent of the source spectrum, ({\it i.e.}, the number of cosmic 
rays starting out at higher energies), but if its existence is confirmed in the
future by more sensitive detectors, it would be evidence for the GZK effect.

Scully and Stecker (2002) have determined the GZK energy, defined as the
energy for a flux decrease of $1/e$, as a function of redshift. At high
redshifts, the target photon density increases by $(1+z)^3$ and both the 
photon and initial cosmic ray energies increase by $(1+z)$. The results
obtained by Scully and Stecker are shown in Figure \ref{parisf3.eps}. 

\begin{figure}
\centerline{\psfig{figure=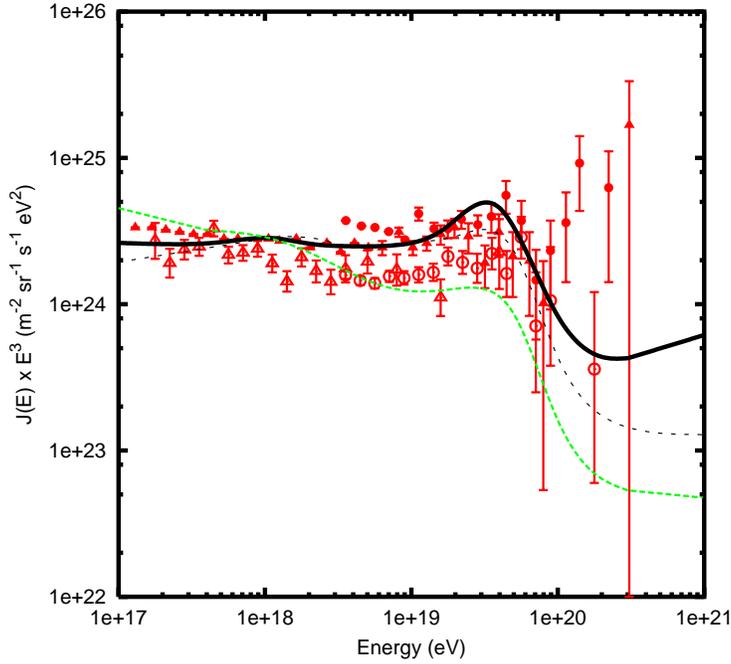,height=14cm}}
\vspace{-1.5cm}
\caption{Predicted spectra for cosmic ray protons as compared with the data.
The middle curve and lowest curve assume an $E^{-2.75}$ source spectrum with
a uniform source distribution and one that follows the $z$ distribution of the
star formation rate respectively. The upper curve is for an $E^{-2.35}$ source
spectrum which requires an order of magnitude more energy input and exhibits
a ``pileup effect''.}
\label{gzkwithdata}
\end{figure}

Figure \ref{gzkwithdata} gives further results of Scully and Stecker
(2002) compared with the present data. It shows the {\it form} of the cosmic 
ray spectrum to be expected from sources with a uniform redshift distribution 
and sources which follow the star formation rate. The required normalization 
and spectral index determine the energy requirements of any cosmological 
sources which are invoked to explain the observations. Pileup effects and GZK 
cutoffs are evident in the theoretical curves in this figure. As can be seen 
in Figure \ref{gzkwithdata}, the present data appear to be 
statistically consistent with either
the presence or absence of a pileup effect. Future data with much better
statistics are required to determine such a spectral structure.

Whereas the {\it AGASA} results indicate a significant number of events at 
trans-GZK energies, the analysis of 
observations made with the {\it HiRes} monocular detector array
show only one event significantly above 100 EeV and appear to be consistent
with the GZK effect (Abu-Zayyad \etal 2002; see Figure \ref{gzkwithdata} and 
section 3.7.)

\subsection{Acceleration and Zevatrons: The ``Bottom Up'' Scenario}

\begin{figure}
\centerline{\psfig{figure=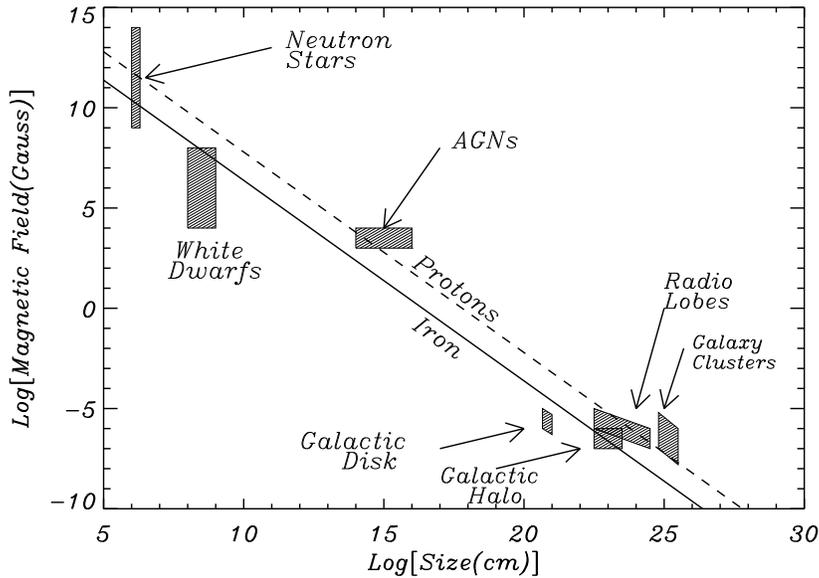,height=8cm}}
\caption{A ``Hillas Plot'' showing potential astrophysical zevatrons 
(from Olinto 2000). The lines are for $B$ {\it vs.} $L$ for
$E_{max}$ = 0.1 ZeV for protons and iron nuclei as indicated.}
\label{Hillas}
\end{figure}

The apparent lack of a GZK cutoff (with the exception of the new {\it HiRes} 
results) has led theorists to go on a hunt for nearby
``zevatrons'', {\it i.e.}, astrophysical sources which can accelerate 
particles to energies $\cal{O}$(1 ZeV = 10$^{21}$eV).

In most theoretical work in cosmic ray astrophysics, it is generally assumed 
that the diffusive shock acceleration process is the most likely mechanism for
accelerating particles to high energy. (See, {\it e.g.}, 
Jones (2000) and references therein.) In this case, the maximum obtainable
energy is given by $E_{max}=keZ(u/c)BL$, where $u \le c$ is the shock speed, 
$eZ$ is the charge of the particle being accelerated, $B$ is the magnetic field
strength, $L$ is the size of the accelerating region and the numerical 
parameter $k = \cal{O}$$(1)$ (Drury 1994). 
Taking $k = 1$ and $u = c$, one finds

$$ E_{max} = 0.9Z(BL) $$

\noindent with $E$ in EeV, $B$ in $\mu$G and $R$ in kpc. This assumes that 
particles can be accelerated efficiently up until the moment when they can 
no longer be contained by the source, {\it i.e.} until their gyroradius 
becomes larger
than the size of the source. Hillas (1984) used this relation to 
construct
a plot of $B$ {\it vs.} $L$ for various candidate astrophysical objects. A
``Hillas plot'' of this kind, recently constructed by Olinto (2000), is shown 
in Figure \ref{Hillas}.
 
Given the relationship between $E_{max}$ and $BL$ as shown in 
Figure \ref{Hillas}., there 
are not too many astrophysical candidates for zevatrons. Of these, galactic
sources such as white dwarfs, neutron stars, pulsars, and magnetars can be 
ruled out because their galactic distribution would lead to anisotropies above
10 EeV which would be similar to those observed at lower energies by
Hayashida {\it et al.} (1999), and this is not the case. Perhaps 
the most promising potential zevatrons are radio lobes of strong radio 
galaxies (Biermann and Strittmatter (1987). The trick is that such 
sources need to be 
found close enough to avoid the GZK cutoff ({\it e.g.}, Elbert and Sommers 
1995) Biermann has further suggested 
that the nearby radio galaxy M87 may be the source of the observed trans-GZK 
cosmic rays (see also Stecker 1968;  Farrar and Piran 2000). Such 
an explanation would require one to invoke magnetic field
configurations capable of producing a quasi-isotropic distribution of 
$> 10^{20}$ eV protons, making this hypothesis questionable. However, if the 
primary particles are nuclei, it is easier to explain a radio galaxy
origin for the two highest energy events (Stecker and Salamon 
1999; see section 3.4). 

\subsubsection{The Dead Quasar Origin Hypothesis}

It has been suggested that since all large galaxies are suspected to
harbor supermassive black holes in their centers which may have once been
quasars, fed by accretion disks which are now used up, that nearby quasar
remnants may be the searched-for zevatrons (Boldt and Ghosh 1999; Boldt
and Lowenstein 2000). This scenario also has potential 
theoretical problems and needs to be explored further. In particular, it
has been shown that black holes which are not accreting plasma cannot
possess a large scale magnetic field with which to accelerate particles
to relativistic energies (Ginzburg and Ozernoi 1964; Krolik 1999;
Jones 2000). 
Observational evidence also indicates that the cores of weakly active
galaxies have low magnetic fields (Falcke 2001 and references therein). 
Another proposed zevatron, the $\gamma$-ray burst, 
is discussed in the next section.

\subsubsection{The Cosmological Gamma-Ray Burst Origin Hypothesis}

In 1995, it was hypothesized that cosmological $\gamma$-ray bursts (GRBs)
could be the zevatron sources of the highest energy cosmic rays 
(Waxman 1995; Vietri 1995).
It was suggested that if these objects emitted the same amount of energy in
ultrahigh energy ($\sim 10^{14}$ MeV) cosmic rays as in $\sim$ MeV photons, 
there would be enough energy input of these particles into intergalactic 
space to account for the observed flux. At that time, it was assumed that the 
GRBs were distributed uniformly, independent of redshift. 

Since 1997, X-ray, optical, and radio afterglows of more than two dozen
GRBs have been detected leading to the subsequent identification of the host
galaxies of these objects and consequently, their redshifts. 
To date, some 27 GRBs afterglows have been detected with a subsequent
identification of their host galaxies.  As of this writing, 26 of the 27 are 
at moderate to high redshifts ($> 0.36$),
with the highest one (GRB000131) lying at a redshift of 4.50.

A good argument in favor of strong redshift evolution for the frequency 
of occurrence of the higher luminosity GRBs has been made by Mao and Mo 
(1998), based on the star-forming nature of the host galaxies. 
The host galaxies of GRBs appear to be sites of active star
formation. The colors and morphological types of the host
galaxies are indicative of ongoing star formation, as
is the detection of Ly$\alpha$ and [OII] in several of these galaxies. 
Further evidence suggests that bursts themselves are directly associated with 
star forming regions within their host galaxies; their positions
correspond to regions having significant hydrogen column densities
with evidence of dust extinction.
It now seems reasonable to assume
that a more appropriate redshift
distribution to take for GRBs is that of the average star
formation rate.
Results of the analysis of Schmidt (1999) also favor a GRB 
redshift distribution which follows the strong redshift evolution of the 
star formation rate. Thus, it now seems reasonable to assume
that a more appropriate redshift
distribution to take for GRBs is that of the average star
formation rate, rather than a uniform distribution.
If we thus assume a redshift distribution for the GRBs which follows the star
formation rate, being significantly higher at higher redshifts, 
GRBs fail by at least an order of magnitude to account
for the observed cosmic rays above 100 EeV (Stecker 2000).
If one wishes to account for the GRBs above 10 EeV, this hypothesis fails by 
two to three orders of magnitude (Scully and Stecker 2002).
Even these numbers are most likely too optimistic, since they are based on the 
questionable assumption of the same amount of GRB
energy being put into ultrahigh energy cosmic rays as in $\sim$ MeV photons.

\subsubsection{Low Luminosity Gamma-Ray Bursts}

An unusual nearby Type Ic supernova, SN 1998bw, at a redshift of 0.0085, 
has been identified as the 
source of a low luminosity burst, GRB980425, with an energy release 
which is orders of magnitude smaller than that for a typical cosmological GRB.
Norris (2002) has given an analysis of the luminosities and
space densities of such nearby low luminosity long-lag GRB sources which
are identified with Type I supernovae. For these sources, he finds a rate
per unit volume of $7.8 \times 10^{-7}$ Mpc$^{-3}$yr$^{-1}$ and an average
(isotropic) energy release per burst of 1.3 $\times 10^{49}$ erg over the
energy range from 10 to 1000 keV. The energy release per unit volume is then
$\sim 10^{43}$ erg Mpc$^{-3}$yr$^{-1}$. This rate is more than an order of
magnitude below the rate needed to account for the cosmic rays with energies
above 10 EeV.

\subsection{The Heavy Nuclei Origin Scenario}

\begin{figure}
\centerline{\psfig{figure=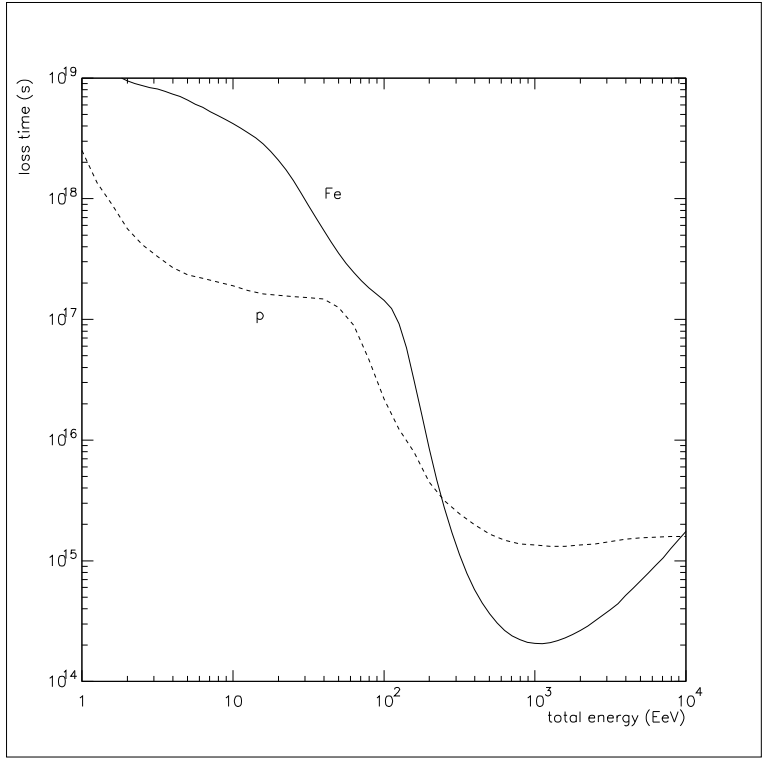,height=13cm}}
\vspace{-1.5cm}
\caption{Mean energy loss times for protons (Stecker 1968; Puget, Stecker and
Bredekamp 1976) and nuclei originating as Fe (Stecker and Salamon 1999).}
\label{parisf6}
\end{figure}

A more conservative hypothesis for explaining the trans-GZK events is that they
were produced by heavy nuclei. Stecker and Salamon (1999) have shown that the
energy loss time for nuclei starting out as Fe is longer than that for protons
for energies up to a total energy of $\sim$300 EeV 
(See Figure \ref{parisf6}.) 

Stanev\ \etal (1995) and Biermann (1998) have examined the arrival 
directions of the highest energy events.
They point out 
that the $\sim 200$ EeV event is within 10$^\circ$ of the direction of the 
strong radio galaxy NGC 315. This galaxy lies at a distance of only 
$\sim$ 60 Mpc from us. For that distance, the results of Stecker and Salamon 
(1999) indicate that heavy nuclei would
have a cutoff energy of $\sim$ 130 EeV, which may be within the uncertainty in
the energy determination for this event. The $\sim$300 EeV event is within
12$^\circ$ of the direction of the 
strong radio galaxy 3C134. The distance to 3C134 
is unfortunately unknown because its location behind a dense molecular cloud 
in our own galaxy obscures the spectral lines required for a measurement of 
its redshift. 

An interesting new clue that we may indeed be seeing heavier nuclei above the
proton-GZK cutoff comes from a recent analysis of inclined air showers
above 10 EeV energy (Ave, {\it et al.} 2000). These new results favor
proton primaries below the p-GZK cutoff energy but they {\it appear to favor a 
heavier composition above the p-GZK cutoff energy}. It will be interesting to
see what future data from much more sensitive detectors will tell us.

\subsection{Top-Down Scenarios: ``Fraggers''}

A way to avoid the problems with finding plausible astrophysical zevatrons is
to start at the top, {\it i.e.}, the energy scale associated with grand
unification, supersymmetric grand unification or its string theory equivalent.

  The modern scenario for the early history of the big bang takes account of 
the work of particle theorists to unify the forces of nature in the framework 
of Grand Unified Theories (GUTs) (\eg , Georgi and Glashow 1974). 
This concept extends the very successful work of Nobel Laureates Glashow, 
Weinberg, and Salam in unifying the electromagnetic and weak nuclear forces 
of nature (Glashow 1960; Weinberg 1967; Salam 1968). 
As a consequence of this theory, the electromagnetic and weak forces would
have been unified at a higher temperature phase in the early history 
of the universe and then would have been broken into separate forces through
the mechanism of spontaneous symmetry breaking caused by vacuum fields which 
are known as Higgs fields.

In GUTs, this same paradigm is used to infer that the electroweak force
becomes unified with the strong nuclear force at very high 
energies of $\sim 10^{24}$ eV 
which occurred only $\sim 10^{-35}$ seconds 
after the big bang. The forces then became separated owing to interactions
with the much heavier mass scale Higgs fields whose symmetry was broken
spontaneously. The supersymmetric GUTs (or SUSY GUTs) provide an explanation
for the vast difference between the two unification scales (known as the
``Hierarchy Problem'') and predict that the running coupling constants 
which describe the strength of the various forces become equal at the SUSY GUT
scale of $\sim 10^{24}$ eV (Dimopoulos, Raby and Wilczek 1982). 

\subsubsection{Topological Defects: Fossils of the Grand Unification Era}

Very heavy ``topological defects" can be produced as a consequence of
the GUT phase transition when the strong and electroweak forces became
separated. These defects are localized regions of vacuum Higgs fields
where extremely high densities of mass-energy are trapped. 
  
Topological defects in the vacuum of
space are caused by misalignments of the heavy Higgs fields in regions which
were causally disconnected in the early history of the universe. These are 
localized regions where extremely high densities of mass-energy are trapped. 
Such defects go by designations such as cosmic strings, monopoles, walls,
necklaces (strings bounded by monopoles), and textures, depending on their 
geometrical and topological properties.  Inside a topological defect 
vestiges of the early universe may be preserved to the present day.  
The general scenario for creating topological defects in the early universe 
was suggested by Kibble (1976).

Superheavy particles or topological structures arising at the GUT energy scale
$M \ge 10^{23}$ eV can decay or annihilate to produce ``$X$-particles'' (GUT 
scale Higgs particles, superheavy fermions, or leptoquark bosons of mass M.) 
In the case of strings
this could involve mechanisms such as intersecting and intercommuting
string segments and cusp evaporation. These $X$-particles will 
decay to produce QCD fragmentation jets at ultrahigh energies, so I will 
refer to them as ``fraggers''. QCD fraggers produce mainly pions, with a 3 to 
10 per cent admixture of
baryons, so that generally one can expect them to produce at least an order of 
magnitude more high energy $\gamma$-rays and neutrinos than protons. 
The same general scenario would hold for the decay of long-lived superheavy 
dark matter particles (see section 3.8), which would also be fraggers. It has 
also been suggested that the decay of ultraheavy particles from topological 
defects produced in SUSY-GUT models which can
have an additional soft symmetry breaking scale at TeV energies (``flat SUSY
theories'') may help explain the observed $\gamma$-ray background flux at
energies $\sim$ 0.1 TeV (Bhattacharjee, Shafi and Stecker 1998).

The number of variations and models for explaining the ultrahigh energy
cosmic rays based on the GUT or SUSY GUT scheme (which have come to be
called ``top-down'' models) 
has grown to be enormous and I will not attempt to list all of the
numerous citations involved. Fortunately, Bhattacharjee and Sigl (2000) 
have published an extensive review with over 500 citations and I 
refer the reader to this review for further details of ``top-down'' models and 
references. The important thing to note here
is that, if the implications of such models are borne out by future cosmic
ray data, they may provide our first real evidence for GUTs.

\subsubsection{``Z-bursts''}

It has been suggested that ultra-ultrahigh energy $\cal{O}$(10 ZeV) neutrinos
can produce ultrahigh energy $Z^0$ fraggers by interactions with  1.9K thermal
CBR neutrinos (Weiler 1982; Fargion\ \etal 1999; Weiler 1999), 
\cite{we99} resulting in ``Z-burst'' fragmentation jets, again producing 
mostly pions. This will occur at the resonance energy $E_{res} =
4[m_{\nu}({\rm eV})]^{-1}$ ZeV. A typical $Z$ boson will decay to produce
$\sim$2 nucleons, $\sim$20 $\gamma$-rays and $\sim$ 50 neutrinos, 2/3 of 
which are $\nu_{\mu}$'s. 

If the nucleons which are produced from Z-bursts originate within a few tens of
Mpc of us they can reach us, even though the original $\sim$ 10 ZeV 
neutrinos could have come from a much further distance. 
It has been suggested that this effect can be amplified if our galaxy has
a halo of neutrinos with a mass of tens of eV (Fargion, Mele and Salis 1999;
Weiler 1999). 
However, a neutrino mass large enough 
to be confined to a galaxy size neutrino halo (Tremaine and Gunn 1979) 
would imply a hot dark matter cosmology which is 
inconsistent with simulations of galaxy formation and clustering 
({\it e.g.}, Ma and Bertschinger 1994) 
and with angular fluctuations in the CBR. (Another problem with halo
fraggers is discussed below in section 3.9) A mixed dark matter model with 
a lighter neutrino mass (Shafi and Stecker 1984) 
produces predicted CBR angular fluctuations (Schaefer, Shafi and Stecker 
1989) which are consistent with the {\it Cosmic Background 
Explorer} data (Wright 1992). In such a model, neutrinos would 
have density fluctuations on the 
scale of superclusters, which would still allow for some amplification 
(Weiler 1999). The tritium decay spectral endpoint
limits on the mass of the electron neutrino (Weinheimer\ \etal 1999), 
together with the very small neutrino flavor mass differences 
indicated by the atmospheric and solar neutrino oscillation results
(Fukuda\ \etal 1998, 1999; Smy 2002; Ahmad\ \etal 2002; Bandyopadhyay\ \etal
2002) constrains all neutrino flavors to have 
masses in the range $\cal{O}$(eV) or less. This is much too small a mass 
for neutrinos to to be confined to halos of individual galaxies.

The basic general problem with the Z-burst explanation
for the trans-GZK events is that one needs to produce 10 ZeV neutrinos. If 
these are secondaries from pion production, this implies that the primary
protons which produce them must have energies of hundreds of ZeV! Since we
know of no astrophysical source which would have the potential of accelerating
particles to energies even an order of magnitude lower (see section 4),
a much more likely scenario for producing 10 ZeV neutrinos would be by a 
top-down process. The production rate of neutrinos from such processes is
constrained by the fact that the related energy release into electromagnetic 
cascades which produce GeV range \grays\ is limited by the satellite 
observations (see the review by Bhattacharjee and Sigl 2000). 
This constraint, together with the low probability for 
Z-burst production, relegates the Z-burst phenomenon to a minor 
secondary role at best.

\subsubsection{Ultraheavy Dark Matter Particles: ``Wimpzillas''}

The homogeneity and flatness of the present universe may imply that a 
period of very rapid expansion,
called inflation, took place shortly after the big bang.  
The early inflationary phase in the history of the universe can 
lead to the production of ultrahigh energy neutrinos. 
The inflation of the
early universe is postulated to be controlled by a putative vacuum field
called the inflaton field. During inflation, the universe is cold but, 
when inflation is over, coherent oscillations of the inflaton field reheat 
it to a high temperature. While the inflaton field is oscillating, non-thermal 
production of very heavy particles (``wimpzillas'') may take place.  
These heavy particles may survive to the present as a part of the dark 
matter. Their decays will produce ultrahigh energy particles and photons 
{\it via} fragmentation.

It has been suggested that such particles may be the source of ultrahigh
energy cosmic rays (Berezinsky \etal 1997; Kuz'min and Rubakov 1998; 
Blasi \etal 2002; Sarkar and Toldr\`{a} 2002; Barbot and Drees 2002). 
The annihilation or decay of such particles in a
dark matter halo of our galaxy would produce ultrahigh energy nucleons
which would not be attenuated at trans-GZK energies owing to their proximity.

\subsubsection{Halo Fraggers and the Missing Photon Problem}

Halo fragger models such as Z-burst and ultraheavy halo dark matter 
(``wimpzilla'') decay or annihilation, as we have seen, will produce
more ultrahigh energy photons than protons. These ultrahigh energy photons 
can reach the
Earth from anywhere in a dark matter galactic halo, because, as shown in
Figure 7, there is a ``mini-window'' for the transmission of ultrahigh 
energy cosmic rays between $\sim 0.1$ and $\sim 10^{6}$ EeV.

Photon-induced giant air showers have an evolution profile which is 
significantly different from nucleon-induced showers because of the
Landau-Pomeranchuk-Migdal (LPM) effect (Landau and Pomeranchuk 1953; Migdal 
1956)  
and because of cascading in the Earth's magnetic field (Cillis\ \etal 1999)
(see Figure 7). By taking this into account, Shinozaki, {\it
et al.} (2002) have used the AGASA data to place upper limits on 
the photon composition of their UHECR showers. They find a photon content 
upper limit of 28\% for events above 10 EeV and 67\% for events above 30 EeV 
at a 95\% confidence level with no indication of photonic showers above 100 
EeV. A recent reanalysis of the ultrahigh energy events observed at Haverah 
Park by Ave, {\it et al.} (2002) 
indicates that less than half of the events (at 95\% confidence level)
observed above 10 and 40 EeV are $\gamma$-ray initiated.
An analysis of the highest energy Fly's Eye event ($E = 300$ EeV) 
shows it not to be of photonic origin,
as indicated in Figure \ref{gprofile}. In addition,
Shinozaki, {\it et al.} (2002) have found no indication of departures from
isotropy as would be expected from halo fragger photonic showers, this
admittedly with only 10 events in their sample.

\begin{figure}
\centerline{\psfig{figure=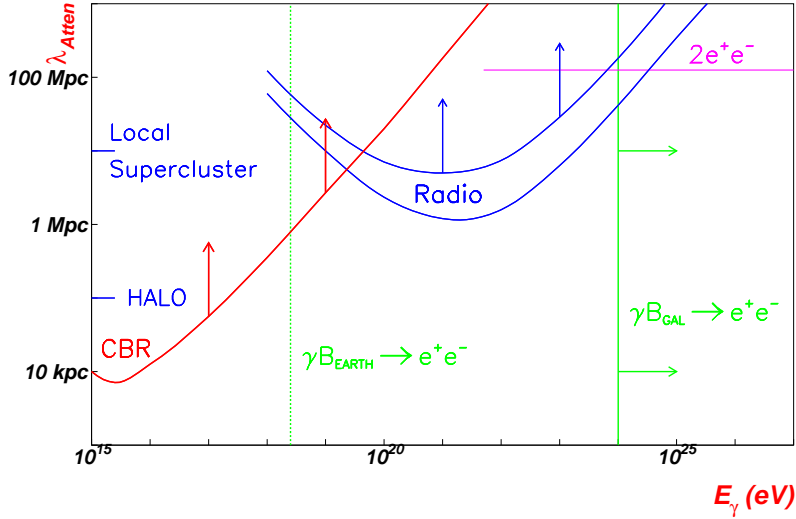,height=8cm}}
\vspace{-0.5cm}
\caption{The mean free path for ultrahigh energy $\gamma$-ray attenuation
{\it vs.} energy. The curve for electron-positron pair production off the 
cosmic background radiation (CBR) is based on Gould and Schr\'{e}der (1966). 
The two estimates for pair production off the extragalactic 
radio background are from Protheroe and Biermann (1996).
The curve for double pair production is based on  Brown, 
{\it et al.} (1973). The physics of pair production by single photons in 
magnetic fields is discussed by Erber (1966). This process
eliminates all photons above $\sim 10^{24}$ eV and produces a terrestrial
anisotropy in the distribution of photon arrival directions above 
$\sim 10^{19}$eV.}
\end{figure}

\begin{figure}
\centerline{\psfig{figure=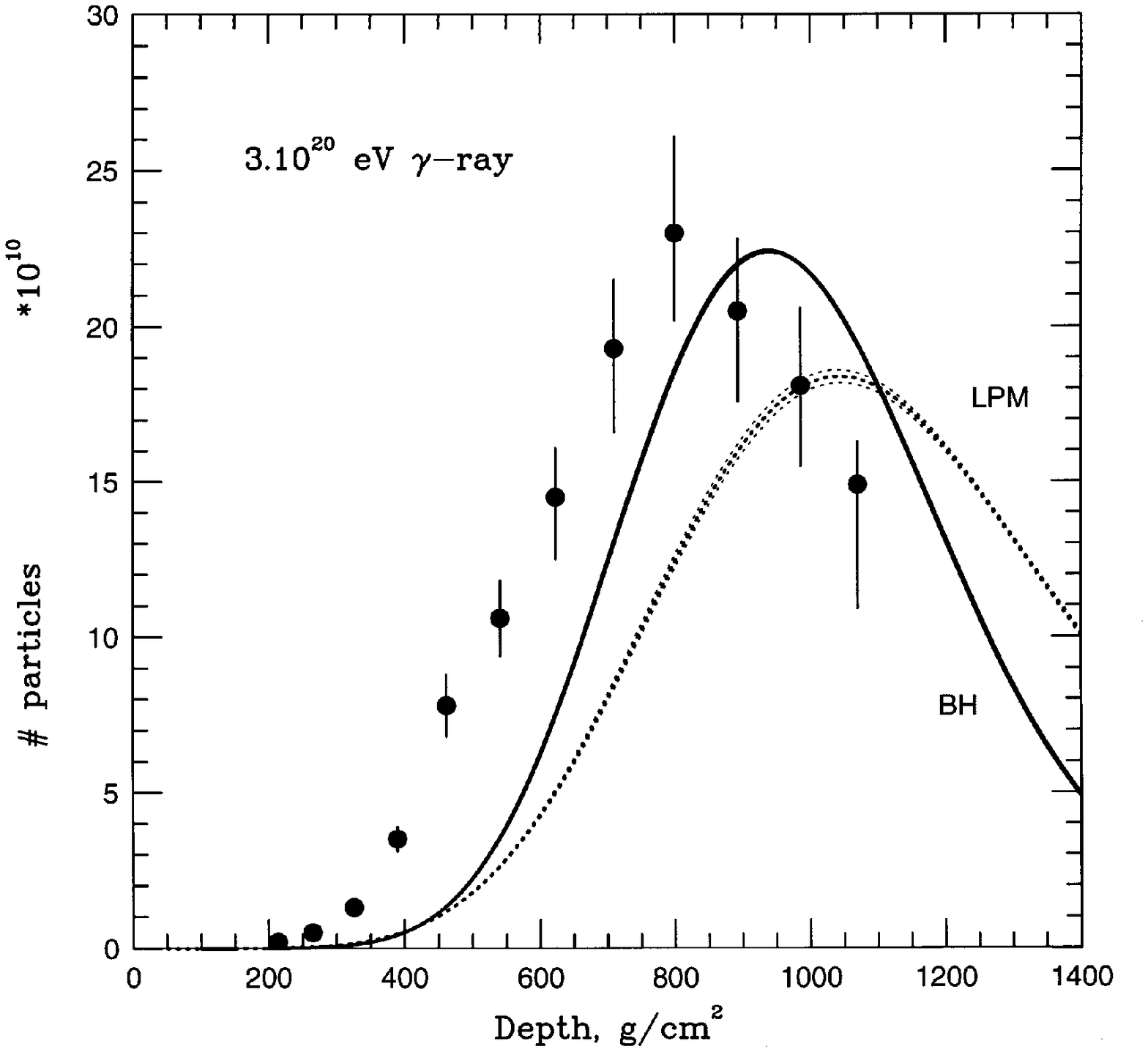,height=7cm}}
\caption{The composite atmospheric shower profile of a 300 EeV photon-induced
shower calculated with the Bethe-Heitler (solid) electromagnetic cross
section and with the LPM effect taken into account (dashed line, see text).
The measured Fly's Eye profile, which fits the profile of a nucleonic 
primary, is shown by the data points (Halzen and Hooper 2002).}
\label{gprofile}
\end{figure}

\subsection{Other New Physics Possibilities}

The GZK cutoff problem has stimulated theorists to look for possible solutions
involving new physics. Some of these involve (A) a large increase in the
neutrino-nucleon cross section at ultrahigh energies, (B) new particles, 
and (C) a small violation of Lorentz Invariance (LI).

\subsubsection{Increasing the Neutrino-Nucleon Cross Section at Ultrahigh 
Energies}

Since neutrinos can travel through the universe without interacting with the
2.7K CBR, it has been suggested that if the neutrino-nucleon cross section 
were to increase to hadronic values at ultrahigh energies, they could produce 
the giant air showers and account for the observations of showers above the
proton-GZK cutoff. Several suggestions have been made for processes that can
enhance the neutrino-nucleon cross section
at ultrahigh energies. These suggestions include composite models of neutrinos
(Domokos and Nussinov 1987; Domokos and Kovesi-Domokos 1988),
scalar leptoquark  resonance channels (Robinett 1988) and the exchange of dual 
gluons (Bordes, {\it et al.} 1998).
Burdman, Halzen and Gandhi (1998)  have ruled out
a fairly general class of these types of models, including those listed above,
by pointing out that in order to increase the neutrino-nucleon cross section
to hadronic values
at $\sim 10^{20}$ eV without violating unitarity bounds, the relevant scale 
of compositeness or particle exchange would have to be of the order of a 
GeV, and that such a scale is ruled out by accelerator experiments.

More recently, the prospect of enhanced neutrino cross sections has been 
explored in the context of extra dimension models. Such models have been
suggested by theorists to unify the forces of physics since the days of
Kaluza (1921) and Klein (1926). In recent years,
they have been invoked by string theorists and by other theorists as a 
possible way for accounting for the extraordinary weakness of the gravitational
force, or, in other words, the extreme size of the Planck mass (Arkani-Hamed,
Dimopoulos and Dvali 1999; Randall and Sundrum 1999). 
These models allow the virtual exchange of gravitons propagating 
in the bulk ({\it i.e.} in the space of full extra dimensions) 
while restricting the propagation of other particles to the familiar four 
dimensional space-time manifold. It has been suggested that in such models,
$\sigma$($\nu$N) $\simeq [E_{\nu}/(10^{20}\rm eV)]$ mb (Nussinov and Schrock
1999; Jain, {\it et al.} 2000; see also Domokos and Kovesi-Domokos 1999). 
It should be noted that a cross section of
$\sim 100$ mb would be necessary to approach obtaining consistency with
the air shower profile data. Other scenarios involve the neutrino-initiated
atmopheric production of black holes (Anchordoqui\ \etal 2002) 
and even higher dimensional extended objects, p-dimensional branes called 
``p-branes'' (Ahn, Cavaglia and Olinto 2002; Anchordoqui, Feng and Goldberg 
2002). 
Such interactions, in principle, can increase the neutrino total atmospheric 
interaction cross section by orders of magnitude above the standard model
value. However, as discussed by Anchordoqui, Feng and Goldberg (2002), 
sub-mm gravity experiments and astrophysical constraints rule
out total neutrino interaction cross sections as large as 100 mb as would be
needed to fit the trans-GZK energy air shower profile data. Nonetheless, extra
dimension models still may lead to significant increases in the neutrino
cross section, resulting in moderately penetrating air showers. Such 
neutrino-induced showers should also be present at somewhat lower energies and
provide an observational test for extra dimension TeV scale gravity models
(Anchordoqui\ \etal 2001; Tyler, Olinto and Sigl 2001). 
As of this writing, no such showers have been observed, putting an indirect 
constraint on fragger scenarios with TeV gravity models.

\subsubsection{New Particles}

The suggestion has also been made that new neutral particles containing
a light gluino could be producing the trans-GZK events (Farrar 1996;
Cheung, Farrar and Kolb 1998).
While the invocation of such new particles
is an intriguing idea, it seems unlikely that such particles of a few
proton masses would be produced in copious enough quantities in astrophysical
objects without being detected in terrestrial accelerators. Also there
are now strong constraints on gluinos (Alavi-Harati, {\it et al.} 1999).
One should note that while it is true that the GZK threshold for such 
particles would be higher than that for protons, 
such is also the case for the more prosaic heavy nuclei
(see section 3.4). In addition, such neutral particles cannot be accelerated 
directly, but must be produced as secondary particles, making the energetics
reqirements more difficult.

\subsubsection{Breaking Lorentz Invariance}  

With the idea of spontaneous symmetry breaking in particle physics came the
suggestion that Lorentz invariance (LI) might be weakly broken at high energies
(Sato and Tati 1972). Although no real quantum theory of gravity exists, it 
was suggested that LI might be broken as a consequence of such a theory
(Amelino-Camilia {\it et al.} 1998). A simpler formulation
for breaking LI by a small first order perturbation in the electromagnetic 
Lagrangian which leads to a renormalizable treatment has been given by
Coleman and Glashow (1999). Using this formalism, these authors 
have shown than only a very tiny amount of LI symmetry breaking is required 
to avoid the GZK effect by supressing photomeson interactions between
ultrahigh energy protons and the CBR. This LI breaking amounts to a 
difference of $\cal{O}$($10^{-23}$) between the maximum proton pion 
velocities. By comparison, Stecker and Glashow (2001) have placed an upper 
limit of $\cal{O}$($10^{-13}$) on the difference between the velocities of the 
electron and photon, ten orders of magnitude higher than required to eliminate
the GZK effect. 

\subsection{Is the GZK Effect All There Is?} 

There is a remaining ``dull'' possibility. Perhaps the GZK effect is  
consistent with the data and is all there is at ultrahigh energies.
The strongest case for trans-GZK physics comes from the {\it AGASA} results.
The {\it AGASA} group, which reported up to 17 events with energy greater than
or equal to $\sim$ 100 EeV (Sasaki\ \etal 2001), 
has now lowered this number to 8 (see footnote 1). 
However, the {\it HiRes} Group have not confirmed the {\it AGASA} results; 
the {\it HiRes} results imply lower fluxes of cosmic rays above $\sim$ 100 
EeV which are consistent with the GZK effect
(Abu-Zayyad\ \etal 2002; see Figure \ref{gzkwithdata}.)
We note, however,  that even if the GZK effect is seen, top-down scenarios 
predict the reemergence of 
a new component at even higher energies (Aharonian, Bhattacharjee and Schramm 
1992; Bhattacharjee and Sigl 2000). 

The {\it AGASA} data indicate a significant deviation from pure GZK even if 
the source number is weighted like the local galaxy distribution (Blanton, 
\etal 2001.) In addition to this discrepency, 
the fact that a flourescence detector, Fly's Eye, reported the 
highest energy event yet seen, {it viz.}, $E \simeq 300$ EeV,
makes the experimental situation interesting
enough to justify both more sensitive future detectors and the exploration
of new physics and astrophysics. In this regard, we note that the {\it Auger}
detector array will use both scintillators and flourscence detectors
(Zas 2001; see section 5.) 
Therefore, combined results from this detector array can help clarify the
present discrepency between the {\it AGASA} and {\it HiRes} results.

\subsection{Ultrahigh Energy Event Signatures}

Future data which will be obtained with new detector arrays and satellites
(see next section) will give us more clues relating to the origin of the
trans-GZK events by distinguishing between the various hypotheses which have
been proposed.

A zevatron origin (``bottom-up'' scenario) will produce air-showers primarily
from primaries which are protons or heavier nuclei, with a much smaller 
number of neutrino-induced showers. The neutrinos will be secondaries from 
the photomeson interactions which produce the GZK effect (Stecker 1973; 1979; 
Engel, Seckel and Stanev 2001 and references therein). 
In addition, zevatron events may cluster near the direction of the sources.

A ``top-down'' (GUT) origin mechanism will not produce any heavier nuclei and
will produce more ultrahigh energy neutrinos
than protons. This was suggested as a signature of
top-down models by Aharonian, Bhatacharjee and Schramm (1992). 
Thus, it will be important to look for the neutrino-induced air
showers which are expected to originate much more deeply in the atmosphere than
proton-induced air showers and are therefore expected to be mostly
horizontal showers. Looking for these events can most easily be done with a
satellite array which scans the atmosphere from above (see Section 4.)

Top-down models also produce more photons than protons.
However, the mean free path of these photons against
pair-production interactions with extragalactic low frequency radio photons
from radio galaxies is only a few Mpc at most (Protheroe and Biermann 1996).
The subsequent electromagnetic cascade and synchrotoron emission 
of the high energy electrons produced in the cascade dumps the energy of these 
particles into much lower energy photons (Wdowczyk, Tkaczyk and Wolfendale 
1972; Stecker 1973).
However, the photon-proton ratio is an effective tool for testing halo 
fragger models (See sect. 3.9.)

Another characteristic which can be used to distinguish between the bottom-up 
and top-down models
is that the latter will produce much harder spectra. If differential
cosmic ray spectra are parametrized to be of the form $F \propto E^{-\Gamma}$,
then for top-down models $\Gamma < 2$, whereas for bottom-up models
$\Gamma \ge 2$. Also, because of the hard source spectrum in the ``top-down'' 
models, they should exhibit both a GZK suppression and a pileup just before the
GZK energy.

If Lorentz invariance breaking is the explanation for the missing GZK effect,
the actual absence of photomeson interactions should result the absence of a
pileup effect as well.

\section{Ultrahigh Energy Cosmic Ray Neutrinos}

Astronomy at the highest energies observed must
be performed by studying neutrinos rather than photons 
because the universe is opaque to photons originating at redshift $z$
at energies above $100(1+z)^{-2}$ 
TeV owing to interactions with the 2.7 K radiation (Stecker 1969)
and even lower energies owing to interactions with other sources of
extragalactic radiation (see section 2.4.) On the other hand,
the cross section for neutrino detection rises with energy (see, \eg 
Gandhi\ \etal 1998) making the detection of neutrinos easier at higher
energies. Although no ultrahigh energy cosmic neutrinos have yet been seen,
there are various potential production mechanisms and sources which may
produce these particles and their study can yield important new information
for physics, cosmology, and astrophysics.

\subsection{Neutrinos from Interactions of Ultrahigh Energy Cosmic Rays
with the 3 K Cosmic Background Radiation}

\begin{figure}
\centerline{\psfig{figure=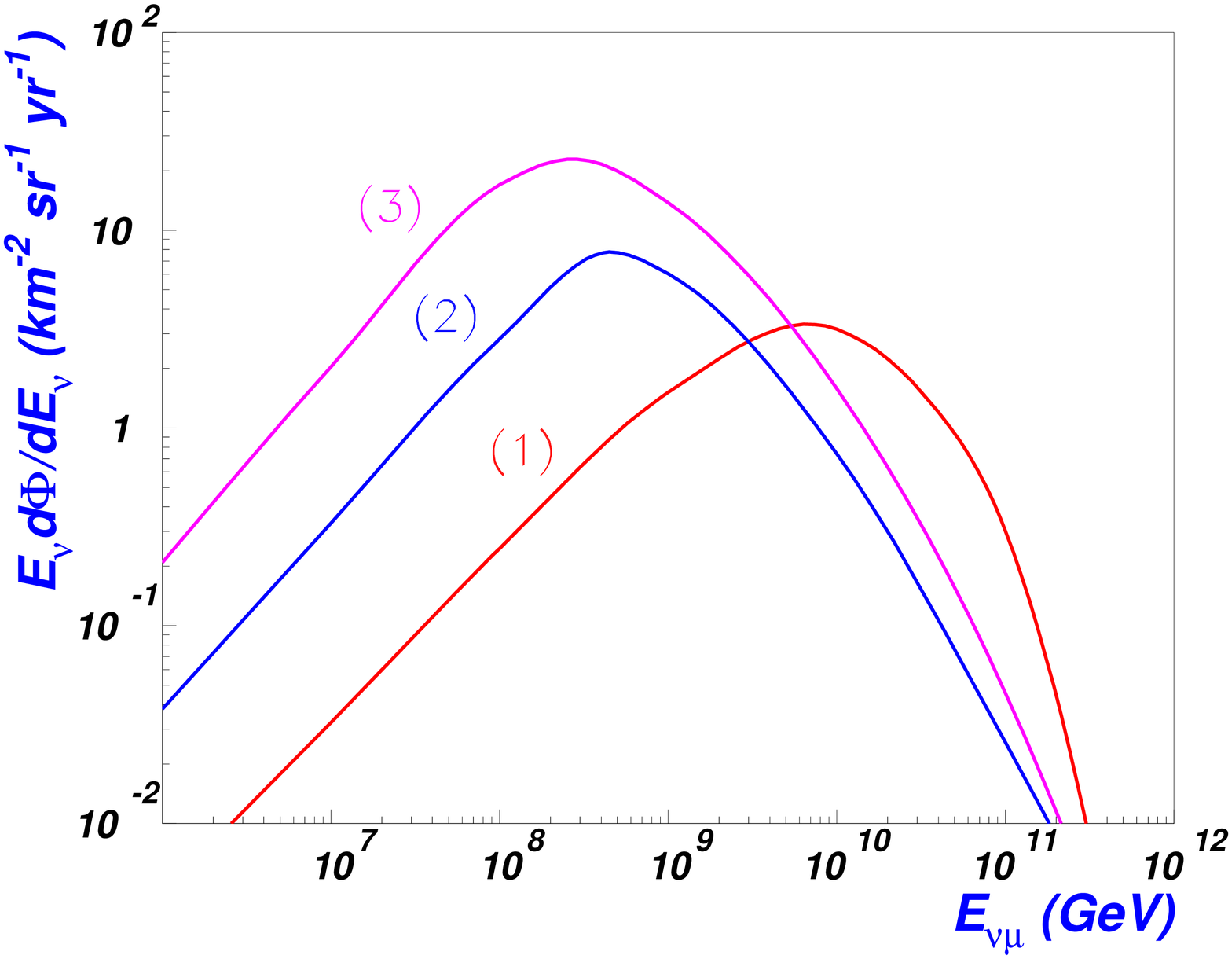,height=12cm}}
\caption{The $\nu_{\mu}$ flux from photomeson production {\it via} 
$p\gamma_{2.7K}$ followed by $\pi^{\pm}$ decay. Curve (1) is the
calclated flux without redshift evolution obtained by Stecker (1979).
The fluxes obtained by Engel \etal (2001) with redshift evolution of the 
proton sources $\propto (1 + z)^m$ with m=3 and 4 respectively are given
by curves (2) and (3).} 
\label{cbrnu}
\end{figure}

Measurements from the {\it COBE} (Cosmic Background Explorer) convincingly 
proved that the universe is filled with radiation having the character of a 
near-perfect 2.725 K black body, which is a
remnant of the big-bang.  As discussed in section 3.2, protons having energies 
above 100 EeV will interact with photons of this radiation, producing pions
(Greisen 1966; Zatsepin and Kuz'min 1966).  
Ultrahigh energy neutrinos will result from decay of these pions 
(Stecker 1973, 1979). The spectrum calculated
by Stecker (1979) without ultrahigh energy source evolution is shown in
Figure \ref{cbrnu} along with two flux spectra assuming source evolution
$\propto (1 + z)^m$ calculated by Engel\ \etal (2001). 

By extrapolating the present measurements of 
the flux of such high energy protons (see section 3), it can be shown that 
measurable numbers of high energy neutrinos can be detected using imaging 
optics aboard satellites looking down at the luminous tracks produced in the 
atmosphere by showers of
charged particles produced when these neutrinos hit the nuclei of 
atoms in the atmosphere (see section 5).

\subsection{Neutrinos from Active Galactic Nuclei}

Quasars and other active galactic nuclei (AGN) are most powerful 
continuous emitters of energy in the known universe.
These remarkable objects are fueled by the gravitational energy released by 
matter falling into a supermassive black hole at the center of the quasar
core. The infalling matter accumulates in an accretion disk which heats
up to temperatures high enough to emit large amounts of UV and soft 
X-radiation. The mechanism responsible for the efficient conversion of 
gravitational energy to observed luminous energy in not yet completely 
understood. If this conversion occurs partly through the acceleration of 
particles to relativistic energies, perhaps by the shock formed at the inner 
edge of the accretion disk (Kazanas and Ellison 1986), then the interactions 
of the resulting 
high energy cosmic rays with the intense photon fields produced by the disk
at the quasar cores can lead to the copious production of mesons. The 
subsequent decay of these mesons will then produce large fluxes of high 
energy neutrinos (Stecker \etal 1991; Stecker and Salamon 1996b). 
Since the \grays\ and high energy cosmic rays deep in the
intense radiation field of the AGN core will lose their energy rapidly and
not leave the source region, these AGN core sources will only be observable
as high energy neutrino sources.

Radio loud quasars contain jets of plasma streaming out from the vicinity of
the black hole, in many cases with relativistic velocities approaching the
speed of light. In a subcategory of quasars, known as blazars, these jets
are pointed almost directly at us with their observed radiation, from radio 
to \gray wavelengths, beamed toward us (See sect. 2.3.) It has been found 
that most of these blazars actually emit the bulk 
of their energy in the high energy \gray\ range. If, as has been suggested, 
the $\gamma$-radiation from these objects is the result of interactions 
of relativistic nuclei (Mannheim and Biermann 1989; Mannheim 1993), 
then high energy
neutrinos will be produced with energy fluxes comparable to the \gray\
fluxes from these objects. On the other hand, if the blazar $\gamma$-radiation 
is produced by purely electromagnetic processes involving only high energy 
electrons, then no neutrino flux will result.
  
\subsection{Neutrinos from Gamma-Ray Bursts (GRBs)} 

GRBs are the most energetic transient phenomenon known in the universe.  
In a very short time of $\sim$ 0.1 to 100 seconds, these bursts 
can release an energy in \grays\ alone of the order 
of $10^{52}$ erg.  They are detected at a rate of about a thousand per year 
by present instruments. It has been proposed that particles can be accelerated in these bursts to energies in excess of $10^{20}$ eV by relativistic shocks
(Waxman 1995; Vietri 1995).

It is now known that most bursts are at cosmological distances corresponding to
moderate redshifts ($z \sim 1$). If cosmic-rays are accelerated in them 
to ultrahigh 
energies, interactions with \grays\ in the sources leading to the production
of pions has been suggested as a mechanism for producing very high energy 
neutrinos as well (Waxman and Bahcall 1997; M\'{e}sz\'{a}ros and Waxman 2001).
These neutrinos would also arrive at the Earth in a burst
coincident with the \gray\ photons.
This is particularly significant since the ultrahigh energy cosmic rays from
moderate redshifts are attenuated by interactions with the 2.7 K microwave 
radiation from the big-bang and are not expected to reach the Earth 
themselves in significant numbers (Stecker 2000).

\subsection{Neutrinos from Topological Defects and Dark Matter}

Topological defects are expected to produce very
heavy particles that decay to produce ultrahigh energy neutrinos.  
The annihilation and decay of these
structures is predicted to produce large numbers of 
neutrinos with energies approaching the predicted energy of grand unification 
(\eg Bhattacharjee and Sigl 2000; see section 3.5.). The discovery of such a
large flux of neutrinos with a hard spectrum and with energies approaching
the energy scale predicted for grand unification would be 
{\it prima facie} evidence for a unified gauge theory of strong and 
electroweak interactions.

An early inflationary phase in the history of the universe can 
also lead to the production of ultrahigh energy neutrinos. 
Non-thermal production of very heavy 
particles (``wimpzillas'') may take place. (See section 3.5.3.)
These heavy particles may survive to the present as a part of dark matter.  
Their decays will produce ultrahigh energy particles and photons {\it via} 
fragmentation, including ultrahigh energy neutrinos (see, \eg Barbot \etal
2002.) 

\subsection{Z-burst Neutrinos}

The observed thermal 2.7 K cosmic background radiation which
permeates the universe as a relic of the big-bang is accompanied by
a 1.96 K cosmic neutrino background of the same thermal big-bang origin
(see \eg Kolb and Turner 1990.)
It has been proposed that high energy neutrinos interacting within
the GZK attenuation distance with the copious 2 K blackbody neutrinos
and annihilating at the Z-boson resonance energy can produce the
observed ``trans-GZK'' air-shower events (Weiler 1982, see section 3.5.2.)
The resulting Z-boson decays to produce a shower of leptons, photons
and hadrons, a.k.a.
a ``Z-burst''. Ultrahigh energy neutrinos are produced by the decay of
the $\pi^{\pm}$'s in the Z-burst. Approximately 50 neutrinos are produced
per burst event; 2/3 of them are $\nu_{\mu}$'s. 
Because the annihilation process is resonant, the Z-burst energy is unique.
It is $E_{Z-burst} = 4m_{\nu}({\rm eV})^{-1}$~ZeV.

The Z-burst hypothesis is based on the assumption that there exists a
significant flux of neutrinos at $E \sim 10$~ZeV, perhaps from 
topological defects. Some predictive consequences of this hypothesis are:
(a) that the direction of the air showers should be close to the directions 
of the sources, (b) that there may be multiple events coming from the 
directions of the strongest sources, and
(c) that there exists a relationship between the 
maximum shower energy attainable and the terrestrially-measured neutrino mass,

As was discussed in section 3.5.2, this production mechanism is quite 
speculative at best.  

\subsection{Neutrino Oscillations and Tau-Neutrino Observatons}

Recent observations of the disappearance of atmospheric $\nu_{\mu}$'s 
relative to $\nu_{e}$'s by the Kamiokande group and also the zenith 
angle distribution of this effect (Fukuda \etal 1998, 1999), 
may be interpreted as evidence of the 
oscillation of this weakly interacting neutrino state (``flavor'') into 
another neutrino flavor, either $\nu_{\tau}$'s  or sterile neutrinos. A 
corollary of such a conclusion is that at least one neutrino state has a 
finite mass. This has very important consequences for our basic theoretical
understanding of the nature of neutrinos and may be evidence for the grand
unification of electromagnetic, weak and strong interactions. It is also 
evidence for physics beyond the ``standard model''. 

If $\nu_{\mu}$'s oscillate into $\nu_{\tau}$'s 
with the parameterization implied by the Super-Kamiokande measurements and
the solar neutrino observations, 
(Ganzalez-Garcia and Nir 2002) then the fluxes of these two
neutrino flavors observed from astrophysical sources should be equal. This is
because cosmic neutrinos arrive from such large distances that many 
oscillations are expected to occur during their journey, equalizing the 
fluxes in both flavor states. Otherwise, the fluxes of 
$\nu_{\tau}$'s from such sources would be much less than those of 
$\nu_{\mu}$'s because $\nu_{\mu}$'s are produced abundantly in the decay of 
pions which are easily produced in cosmic sources, whereas $\nu_{\tau}$'s are 
not.

Upward-moving atmospheric showers induced by $\sim$100 TeV 
and traced back to the direction of a cosmic source such as an AGN or a GRB 
at a distance of 1 Gpc would occur even if the difference of the squares of 
the oscillating mass states were small as $\sim 10^{-17}$ eV$^2$.
Thus, the search for upward moving showers from cosmic $\nu_{\tau}$'s, which 
can propagate thorugh the Earth through regeneration at energies above 
$10^{14}$ eV (Halzen and Saltzberg 1998; see section 5.2), provides a 
test for cosmic high energy $\nu_{\tau}$'s from neutrino oscillations.

Another important signature of ultrahigh energy $\nu_{\tau}$'s 
is the ``double bang'' which they would produce. The first shower is 
produced by the original interaction which creates a $\tau$ lepton and a 
hadronic shower. This is followed by the decay of the 
$\tau$ which produces the second shower bang (Learned and Pakvasa 1995). 
The two bangs are separated by a distance of $\sim$ 91.4 $\mu$m 
times the Lorentz factor of the $\tau$.

\subsection{Neutrino Flux Predictions}

\begin{figure}
\vspace{-1.5cm}
\begin{flushleft}
\mbox{\psfig{file=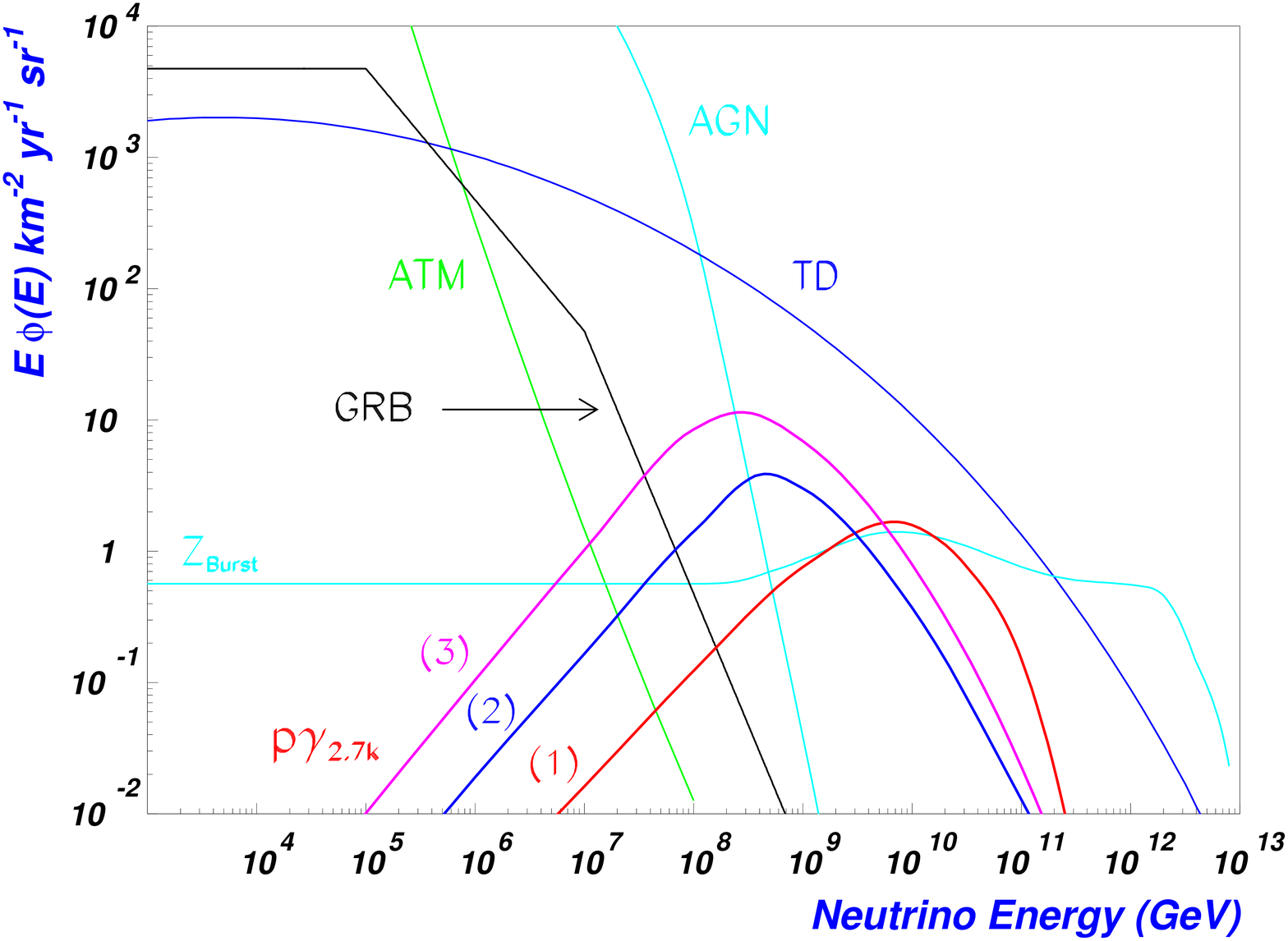,height=4.0in}}
\end{flushleft}
\vspace{-1.0cm}
\caption{Some examples of neutrino flux predictions with $\nu_{\mu}
\leftrightarrow \nu_{\tau}$ oscillations taken into account: 
Atmospheric and AGN fluxes (Stecker and Salamon 1996b); 
Photomeson production {\it via} $p\gamma_{2.7K}$ (as in Figure 
\protect\ref{cbrnu}); 
Topological defects (Sigl, {\it et al.} 1998; 
m$_{X} = 10^{16}$ GeV supersymmetric model); $Z_{Bursts}$ (Yoshida, Sigl, 
and Lee 1998, m$_{\nu}$ = 1 eV, primary flux $\propto E^{-1}$);
$\gamma$-ray bursts (Waxman and Bahcall 1997).}
\label{nuflux}

\end{figure}

Figure \ref{nuflux} illustrates some high energy neutrino flux predictions from
various astrophysical sources discussed above as a function of neutrino 
energy.  Note that the curves show the differential 
$\nu_{\mu} + \bar{\nu_{\mu}}$ flux 
multiplied by $E_{\nu}$. In this figure, $\nu_{\mu} \leftrightarrow
\nu_{\tau}$ oscillations are assumed to reduce the cosmic high energy
neutrino fluxes (including those shown in Figure \ref{cbrnu}) by a factor
of 2.

In the energy range of 0.1 to 
100 PeV, the AGN neutrino flux may dominate over other 
sources. However, neutrinos from individual \gray\ bursts may be 
observable and distinguishable {\it via} their directionality and 
short, intense time characteristics. 
The time-averaged background flux from all bursts is shown in the figure. 

In the energy range $E \ge$ 1 EeV, neutrinos are produced from photomeson
interactions of ultrahigh enrgy cosmic rays with the 2.7 K background photons.
The highest energy neutrinos ($E \ge 100$
EeV) are presumed to arise from more the speculative physics of topological 
defects and Z-bursts.

\subsection{Observational Signatures and Rates}

\subsubsection{Signatures}

The proposed high energy neutrino sources also have different signatures in
terms of other observables which include coincidences with other
observations (GRB's), anisotropy (Z-burst's), and specific relations
to the number of hadronic or photonic air showers also induced
by the phenomena (topological defects, Z-burst's, and 2.7K photomeson 
neutrinos). 

The distiguishing characteristics of astrophysical neutrino
sources are summarized in the Table 1. In the table, the characteristic 
neutrino energy for photomeson 
processes is defined as $\sim 1.8 \times 10^{-2} M_{p}^2/\bar{\epsilon}$ where 
$\bar{\epsilon}$ is the mean energy of the photon field (Stecker 1979),
except for the case of GRB neutrinos where the energy is boosted by a
factor of $\Gamma^2$ where $\Gamma \sim 10^3$ is the bulk Lorentz factor of
the GRB fireball (Waxman and Bahcall 1997).

\begin{table}[h]
	\vspace{0.5cm}
\centering
\begin{tabular}{llllll}
 \hline
 Test & GRB & AGN & TD & Z-Burst & p$\gamma_{2.7K}$ \\
 \hline
 Coincidence & & & & & \\
 with a GRB   & X & - & - & - & - \\
 & & & & & \\
 $(N_{\nu}/N_{p}) \gg 1$& - & - & X & X & - \\
 & & & & & \\
 $(N_{\gamma}/N_{p}) \gg 1$& - & - & X & X & - \\
 & & & & & \\
 Anisotropy & - & -  & - & X & - \\
 & & & & & \\
 Characterisitic & & & & & \\
 Energy & $10^{16}$ eV & $10^{14}$ eV & $10^{21}$ eV & {\Large $\frac{€10^{20}~{\rm 
 eV}}{m_{\nu} ~({\rm eV})}$} & $10^{19}$ eV \\
 & & & & & \\
 Multiple & & & & & \\
 Events & X & X & - & X & - \\
 \hline 
\end{tabular}
\caption{Distinguishing characteristics of the different sources of 
ultra-high energy neutrinos (revised from Cline and Stecker 2000.)}
\end{table}

The distribution of the atmospheric depth of neutrino interactions is 
approximately uniform due to the extremely long interaction path of neutrinos 
in the atmosphere.
This offers a unique signature of neutrino-induced airshowers as a 
significant portion of the neutrino interactions will be deep in the 
atmosphere, i.e. near horizontal, and well separated from airshowers 
induced by hadrons and photons.  

At ultrahigh energies, the cross sections for $\nu$ and $\bar{\nu}$
interactions with quarks become equal (Gandhi\ \etal\ 1998).
The interactions produce an ultrahigh energy lepton which carries off
$\sim$ 80\% of
the incident neutrino energy. The remaining 20\% is
in the form of a hadronic cascade.
Charged current neutrino interactions will, on average, yield an UHE 
charged lepton and a hadronic airshower. At these energies,
electrons will generate electronic airshowers while
$\mu$'s and $\tau$'s will produce airshowers with reduced
particle densities and thus, reduced fluorescence signals.  
As discussed above, he $\nu_{\tau}$'-induced showers have a
``double-bang'' signature. For example, a 10 EeV, $\tau$ decays after traveling
$\gamma c \tau = 500$ km, producing a second airshower which is very well
separated from the first, hadronic airshower at the neutrino interaction point.

\section{Present and Future Detectors}

Of the ground-based ultrahigh energy arrays, the {\it AGASA} array of particle
detectors in Japan is continuing to obtain data on ultrahigh energy cosmic 
ray-induced air showers. Its aperture is 200 km$^2$sr. 
The {\it HiRes} array is an extension of the Fly's Eye which pioneered the 
technique of measuring the 
atmospheric fluorescence light in the near UV (300 - 400 nm range). This light
is isotropically emitted by nitrogen molecules that are excited by the 
charged shower secondaries at the rate of $\sim$4 photons per meter per 
particle. The estimated aperture of the {\it HiRes} monocular detector is 
$\sim$1000 km$^2$sr at 100 EeV after inclusion
of a 10\% duty cycle (Sokolsky 1998). 

The southern hemisphere {\it Auger} array is expected to be on line in the 
near future. This will be a hybrid array which will consist of 1600 particle
detector elements similar to those at Havera Park and three or four 
flourescence detectors (Zas 2001). Its expected aperture will be 7000 
km$^2$sr for the
ground array above 10 EeV and $\sim$ 10\% of this number for the hybrid array.

The next big step will be to orbit a system of space-based detectors which
will look down on the Earth's atmosphere to detect the trails of nitrogen
flourescence light made by giant extensive air showers.
The Orbiting Wide-angle Light collectors ({\it OWL}) 
mission is being proposed 
to study such showers from satellite-based platforms in low 
Earth orbit (600 - 1200 km). {\it OWL} would observe extended air showers 
from space via the air fluorescence technique, thus determining the 
composition, energy, and arrival angle 
distributions of the primary particles in order to deduce their origin.  
Operating from space with a wide field-of-view instrument
dramatically increases the observed target volume, and consequently the 
detected air shower event 
rate, in comparison to ground based experiments. The {\it OWL} baseline 
configuration will yield event rates that are more than two orders 
of magnitude larger than currently operating ground-based experiments.  
The estimated aperture for a two-satellite system is $2.5 \times 10^5$
km$^2$sr above a few tens of EeV after assuming a 10\% duty cycle.

{\it OWL} will be capable of
making accurate measurements of giant air shower events with high statistics. 
It may be able to detect $\cal{O}$(1000) giant air showers per year with 
$E \ge $ 100 EeV (assuming an extrapolation of the
cosmic ray spectrum based upon the AGASA data). 
The European Space Agency is now studying the feasibility of placing such a 
light collecting detector on the International Space Station in order to 
develop the required technology to observe the flourescent trails of giant 
extensive air showers,
and to serve as a pathfinder mission for a later free flyer. This experiment
has been dubbed {\it EUSO} (Extreme Universe Space Observatory; see paper
of Livio Scarsi, these proceedings.)
Owing to the orbit parameters and constraints of the International
Space Station, the effective aperture for EUSO will not be as large 
as that of a free flyer mission. 
A recent compendium of papers on observing giant air showers from space may be
found in Krizmanic, Ormes and Streitmatter (1998).

\subsection{Detection of Neutrino Events from Space}

The proposed Orbiting Wide-angle Light collectors satellite mission, 
{\it OWL}, would have the unmatched capacity to map the arrival directions of
cosmic rays over the entire sky and thus to reveal the locations of strong 
nearby sources and large-scale anisotropies, this owing to
the magnetic stiffness of charged particles of such high energy.  
Thus, with such a detector system, one can investigate 
energy spectra of any detected sources and also time correlations 
with high energy $\nu$'s and \grays.

Preliminary Monte Carlo simulations for an {\it OWL} 
space-based detector (J. Krismanic, private communication)
have indicated that charged current electron neutrino interactions
can be identified with a neutrino aperture
of 20 km$^{2}$-ster at a threshold
energy of 30 EeV and the aperture size
grows with energy $\propto E_{\nu}^{0.363}$ with the
increase in neutrino
cross section (Gandhi \etal 1998).   Event rates can be obtained by convolving
this neutrino aperture with neutrino flux predictions and
integrating.  Note that the neutrino interaction
cross section is included in the definition of neutrino aperture.
Assuming a 10\% duty cycle of the experiment,
The $\nu_{e}$ event rates from several possible UHE neutrino sources
as shown in Figure \ref{nuflux} are found to be as follows: neutrinos from 
the interaction of UHE protons with the microwave background 
(p$\gamma_{2.7K}$)----5 events per year; topological defects----10 
events per year; neutrinos from $Z$-bursts----9 events per year.

\subsection{Upward \v Cerenkov Events from Cosmic Tau-Neutrinos}

The ensemble of charged particles in an airshower will 
produce a large
photon signal from \v Cerenkov radiation which is strongly peaked in the 
forward direction and which is much stronger than the signal
due to air fluorescence at a given energy.  This
translates into a much reduced energy threshold for observing
airshowers {\it via} \v Cerenkov radiation.  As this signal is
highly directional, an orbitting instrument will
only observe those events where the airshower is moving towards
the experiment with the instrument located in the field of the
narrow, \v Cerenkov cone. Thus, it is possible ot observe upward
moving events from $\nu_{\tau}$'s which have propagated through the Earth
(see section 4.5.)

Virtually all particles, including neutrinos with $E \ge 40$ TeV, 
are attenuated by the Earth. However, $\nu_{\tau}$'s will regenerate 
themselves, albeit at a lower energy, due to the
fact that both charged and neutral current interactions will have
a $\nu_{\tau}$ in the eventual, final state (Halzen and Saltzberg 1998).  
The $\nu_{\tau}$ interactions produce a leading high-energy $\tau$ which
then decays to produce another $\nu_{\tau}$. 

For $\nu_{\tau}$ interactions in the Earth's crust, 
the $\tau$'s will have a flight path of length $\gamma c \tau 
~(\approx 50~{\rm m}$ at 1 PeV before they decay.
Those events which occur
at a depth less than $\gamma c \tau$ will produce a $\tau$ coming
out of the Earth and generating an airshower.  For a target
area of 10$^{6}$ km$^{2}$, this yields a target mass
of $10^{8}(E_{\nu}$(GeV) metric tons, \eg
$10^{14}$ metric tons at an energy of 1 PeV for the Earth as
an effective detector mass.

Preliminary investigations of the response of an {\it OWL} space-based 
detector have indicated that the experiment would have
a threshold energy of about 0.1 PeV to upward, Cherenkov airshowers.
Assuming that the Super-Kamiokande atmospheric neutrino
results are due to
$\nu_{\mu} \leftrightarrow \nu_{\tau}$ oscillations,  the 
predicted AGN $\nu_{\mu}$ flux (Stecker and Salamon 1996b)
indicates that {\it OWL} could observe several hundred $\nu_{\tau}$
events per year.  Thus {\it OWL} would be able measure the flux
of putative AGN neutrinos and observe their oscillations into $\nu_{\tau}$'s.

\section*{Acknowledments}
I would like to thank J. Krizmanic for his help in generating some of the
figures and for supplying information particularly for section 5. I would 
also like to thank S. Scully for generating two of the figures and G.
Amelino-Camelia for e-mail discussions.

{}


\begin{thebibliography}{999}

\bibitem[]{ab00} Abu-Zayyad, T. {\it et al.} (2000), \prl {\bf 84}, 4276.
\bibitem[]{ab02} Abu-Zayyad, T. {\it et al.} (2002), e-print astro-ph/0208243.
\bibitem[]{ah92} Aharonian, F., Bhattacharjee, P. and Schramm, D.N. (1992),
{\it Phys. Rev. D} {\bf 46}, 4188.
\bibitem[]{ah98} Aharonian, F.A \etal\ (1997), \aa\ {\bf 327}, L5.
\bibitem[]{ah99} Aharonian, F.A. \etal\ (1999), \aa\ {\bf 349}, 11.
\bibitem[]{ah01a} Aharonian, F.A. \etal\ (2001a), \aa\ {\bf 366}, 62.
\bibitem[]{ah01b} Aharonian, F.A. \etal\ (2001b), \aa {\bf 384}, 834.
\bibitem[]{ah02} Aharonian, F.A. \etal (2002), \epr 0202072.
\bibitem[]{ah02} Ahmad, Q.R.\ \etal (2002), \prl\ {\bf 89}, 011302.
\bibitem[]{an02} Ahn, E.-J., Cavaglia, M. and Olinto, A.V. 2002,
e-print hep-th/0201042.
\bibitem[]{al99} Alavi-Harati, A. {\it et al.} (1999), \prl\ {\bf 83}, 2128.
\bibitem[]{al02} Alvarez-M\~{u}niz, J. and Halzen, F. (2002), \apj , in press,
e-print astro-ph/0205408.
\bibitem[]{am02} Amelino-Camilia, G. (2002), e-print gr-qc/0212002.
\bibitem[]{am98} Amelino-Camilia, G. {\it et al.}, {\it Nature} {\bf 393}, 763.
\bibitem[]{afg02} Anchordoqui, L.A., Feng, J.L. and Goldberg, H. (2002),
{\it Phys. Lett. B} {\bf 535}, 302.
\bibitem[]{an01} Anchordoqui, L.A.\ \etal (2001), {\it Phys. Rev. D} {\bf
63}, 124009.
\bibitem[]{an02} Anchordoqui, L.A.\ \etal (2002), {\it Phys. Rev. D} {\bf
65}, 124027.
\bibitem[]{ar99} Arkani-Hamed, N. Dimopoulos, S. and Dvali, G.R. (1999), {\it 
Phys. Rev. D} {\bf 59}, 086004.
\bibitem[]{at02} Atkins, R.\ \etal\ (2002), \epr/0207149.
\bibitem[]{av00} Ave, M. {\it et al.} (2000), \prl\ {\bf 85}, 2244.
\bibitem[]{av02} Ave, M. {\it et al.} (2002), {\it Phys. Rev.} {\bf 65},
063007.
\bibitem[]{ba34}Baade, W. and Zwicky, F. (1934), 
{\it Phys. Rev.} {\bf 45}, 138.
\bibitem[]{ba02} Bandyopadhyay, A. \etal (2002), {\it Phys. Rev.} {\bf 65}, 
073031.
\bibitem[]{bd02} Barbot, C. and Drees, M. (2002), {\it Phys. Letters B}
{\bf 533}, 107.
\bibitem[]{bar02} Barbot, C. \etal (2002), e-print hep-ph/0205230.
\bibitem[]{be88} Berezinsky, V.S. and Grigor'eva S.I. (1988), 
\aa\ {\bf 199}, 1.
\bibitem[]{be97} Berezinsky, V., Kachelriess, M. and Vilenkin (1997),
\prl {\bf 79},4302.
\bibitem[]{be98} Bershady, M.A., Lowenthal, J,D. and Koo, D.C. (1998), \apj, 
505, 50.
\bibitem[]{bh98} Bhattacharjee, P., Shafi, Q. and Stecker, F.W. (1998), \prl 
{\bf80}, 3698.
\bibitem[]{bh00} Bhattacharjee, P. and Sigl, G. (2000), {\it Phys. Rpts.}
{\bf327}, 109.
\bibitem[]{bi98} Biermann, P.L. (1998) in
{\it Workshop on Observing Giant Cosmic Ray Air Showers from Space},
ed. J.F. Krizmanic, J.F. Ormes  and R.E. Streitmatter, (New York: American 
Institute of Physics, AIP CP-433) p.22. 
\bibitem[]{bi87} Biermann, P.L. and Strittmatter, P.A. (1987), \apj {\bf 322}, 
643.
\bibitem[]{bil98} Biller, S. \etal (1998), \prl {\bf 80}, 2992.
\bibitem[]{bi93} Bird, D.J. {\it et al.} (1993), \prl {\bf 71}, 3401.
\bibitem[]{bi94} Bird, D.J. {\it et al.} (1994), \apj {\bf 424}, 491.
\bibitem[]{bl01} Blanton, M., Blasi, P. and Olinto, A.V. (2001), {\it 
Astropart. Phys.} {\bf 15}, 275.
\bibitem[]{bl02} Blasi, P., Dick, R. and Kolb, E.W. (2002), {\it Astroparticle
Physics} {\bf 18}, 57.
\bibitem[]{bl96} Bloom, E.D. (1996), {\it Space Sci. Rev.} {\bf 75}, 109.
\bibitem[]{bo99} Boldt, E. and Ghosh, P. (1999) \mnras {\bf 307}, 491.
\bibitem[]{bo00} Boldt, E. and Lowenstein, M. (2000), \mnras {\bf 316}, L29.
\bibitem[]{bo98} Bordes, J. {\it et al.} (1998), {\it Astropart. Phys.}
{\bf 8}, 135.
\bibitem[]{br73} Brown, R.W. {\it et al.} (1973), {\it Phys. Rev. D} 
{\bf 8}, 3083.
\bibitem[]{bu00} Bulik, T. \etal (2000), {\it Mon. Not. Royal Astr. Soc.} 
{\bf 31}, 97.
\bibitem[]{bu98} Burdman, G., Halzen, F. and Ghandi, R. (1998), {\it Phys. 
Letters B} {\bf 417}, 107.
\bibitem[]{bu02}Butt, Y.M. \etal (2002), {\it Nature} {\bf 418}, 499.
\bibitem[]{ca97} Catanese, M., Bradbury, S.M., Breslin, A.C., {\it et al.}, 
(1997), \apj {\bf 487}, L143
\bibitem[]{ca98} Catanese, M. \etal\ (1998), \apj\ {\bf 501}, 616.
\bibitem[]{chad98} Chadwick, P.M. \etal\ (1999), {\it Astropart. Phys.} 
{\bf 11}, 145.
\bibitem[]{ch84} Cheng, K.S., Ho, C. and Ruderman, M.A. (1984), 
\apj {\bf 300}, 500.
\bibitem[]{ch98} Cheung, D.J.H., Farrar, G.R. and Kolb, E.W. (1998), 
{\it Phys. Rev. D} {\bf 57}, 4606.
\bibitem[]{ci99} Cillis, A.N. \etal (1992), {\it Phys. Rev. D} {\bf 59}, 
113012.
\bibitem[]{cl00} Cline, D. and Stecker, F.W. (2000), {\it 1999 UCLA 
Workshop on 
Ultrahigh Energy Neutrino Astrophysics Science White Paper}, astro-ph/0003459.
\bibitem[]{co99} Coleman, S. and Glashow, S.L. (1999), Phys. Rev. {\bf D59}, 
116008.
\bibitem[]{co98} Colladay, D. and Kostelecky, V.A. (1998), 
{\it Phys. Rev. D} {\bf 58}, 116008. 
\bibitem[]{da99} Dai \etal (1999), \apj {\bf 511}, 739.
\bibitem[]{da96} Daugerty, J.K. and Harding, A.K. (1996), \apj {\bf 458}, 278.
\bibitem[]{de92} De Jager, O.C. and Harding, A.K. (1992), \apj\ {\bf 396}, 161.
\bibitem[]{de02} De Jager, O.C. and Stecker, F.W. (2002), \apj {\bf 566}, 738 
(DS02).
\bibitem[]{di82} Dimopoulos, S. Raby, S. and Wilczek, F. (1982), {\it Physics
Letters B} {\bf 112}, 133.
\bibitem[]{dj02} Djannati-Atai\ \etal (2002), \epr 0207618.
\bibitem[]{do88} Domokos, G. and Kovesi-Domokos, S. (1988) 
{\it Phys. Rev. D} {\bf 38}, 2833.
\bibitem[]{do99} Domokos, G. and Kovesi-Domokos, S. (1999) \prl {\bf 82}, 1366.
\bibitem[]{do87} Domokos, G. and Nussinov, S. (1987) {\it Phys. Letters B} 
{\bf 187}, 372.
\bibitem[]{dr94} Drury, L. (1994) {\it Contemp. Phys.} {\bf35},232.
\bibitem[]{dav94} Drury, L., Aharonian, F.A. and V\"{o}lk, H.J. (1994), \aa\ 
{\bf 287}, 959. 
\bibitem[]{dw01} Dwek, E. (2001), in {\it The Extragalactic Infrared 
Background and its Cosmological Implications, IAU Symposium no. 204}, ed. M. 
Harwit and M.G. Hauser, p. 389.
\bibitem[]{dw98a} Dwek, E and Arendt, R.G. (1998), \apj {\bf 508}, L9.
\bibitem[]{dw98b} Dwek, E. {\it et al.} (1998), \apj {\bf 508}, 106.
\bibitem[]{el99} Elbaz, D. {\it et al.}, 
(1999), \aa\ {\bf 351}, L37.
\bibitem[]{el95} Elbert, J.W. and Sommers, P. (1995), \apj {\bf 441}, 151.
\bibitem[]{en01} Engel, R., Seckel, D. and Stanev, T. (2001), {\it Phys.
Rev. D} {\bf 64}, 093010.
\bibitem[]{en02} Enomomto, R. \etal\ (2002), {\it Nature} {\bf 416}, 823.
\bibitem[]{er66} Erber, T. (1966), {\it Rev. Mod. Phys.} {\bf 38}, 626.
\bibitem[]{fa01} Falcke, H. (2001), {\it Rev. Mod. Astron.} {\bf 14}
15.
\bibitem[]{fall96} Fall, S.M., Charlot, S. and Pei, Y.C. (1996),
\apj\ {\bf 464}, L43.
\bibitem[]{fa99} Fargion, D., Mele, B. and Salis, A. (1999), \apj {\bf 517}, 
725.
\bibitem[]{fa96} Farrar, G.R. (1996), \prl {\bf 76}, 4111.
\bibitem[]{fa00} Farrar, G.R. and Piran, T. (2000), \prl {\bf 84}, 3527.
\bibitem[]{fi97} Fixsen, {\it et al.} (1997), \apj {\bf 490}, 482.
\bibitem[]{fi98} Fixsen, D.J.\ \etal (1998), \apj {\bf 508}, 123.
\bibitem[]{fo00} Fossati, G. \etal (2000), \apj {\bf 541}, 166.
\bibitem[]{fu98} Fukuda\ \etal (1998), \prl 81, 1562.
\bibitem[]{fu99} Fukuda\ \etal (1999), \prl 82, 1810.
\bibitem[]{ga00} Gaisser, T.K. (2000), in {\it Observing Ultrahigh Energy
Cosmic Rays from Space and Earth} CP566 (2000): American Inst. of Physics), 
pg. 566, e-print astro-ph/0011525.
\bibitem[]{ga98} Gaisser, T.K., Protheroe, R.J. and Stanev, T. (1998), \apj 
{\bf 492}, 219.
\bibitem[]{ge74} Georgi, H. and Glashow, S.L. (1974), \prl {\bf 32} 438.
\bibitem[]{gan98} Gandhi, R. {\it et al.} (1998), {\it Phys Rev D} {\bf 58}, 
093009.
\bibitem[]{gi64} Ginzburg, V.L. and Ozernoi, L.M. (1964), {\it Zh. Eksp.
Teor. Fiz.} {\bf 47}, 1030.
\bibitem[]{gi00} Gispert, R., Lagache, G., and Puget, J.-L. (2000),
\aa\ {\bf 360}, 1.
\bibitem[]{gl60} Glashow, S.L. (1960), {\it Nucl. Phys.} {\bf 22}, 579.
\bibitem[]{go02} Gonzalez-Garcia, M.C. and Nir, Y. (2002), {\it Rev. Mod. 
Phys.}, in press, e-print hep-ph/0202058.
\bibitem[]{go66} Gould, R,J. and Schr\'{e}der, G.P. (1966), \prl {\bf 16}, 252.
\bibitem[]{gr66} Greisen, K. (1966) \prl {\bf 16}, 748.
\bibitem[]{gu00} Guy, J.\ \etal (2000), \aa\ {\bf 359}, 419.
\bibitem[]{ha02} Halzen, F. and Hooper, D. (2002), e-print astro-ph/0204527.
\bibitem[]{ha98} Halzen, F. and Saltzberg, D. (1998), {\it Phys. Rev. Letters}
{\bf 81}, 4305.
\bibitem[]{3egret} Hartman, R.C.\ \etal\ (1999), \apj\ {\it Suppl.} {\bf 123}, 
79.
\bibitem[]{hau98} Hauser, M.G., {\it et al.}, (1998), \apj {\bf 508}, 25.
\bibitem[]{hau01} Hauser, M.G. and Dwek, E., (2001), {\it Ann. Rev.} \aa\ 
{\bf39}, 249.
\bibitem[]{ha99} Hayashida, N. {\it et al.} (1999), {\it Astropart. Phys.}
{\bf 10}, 303.
\bibitem[]{hi85} Hill, C.T. and Schramm, D.N. (1985), {\it Phys. Rev. D}
{\bf 31}, 564.
\bibitem[]{hi84} Hillas, A.M. (1984), {\it Ann. Rev. of Astron. and 
Astrophys.} {\bf 22}, 425.
\bibitem[]{hu94} Hurley, K. \etal\ (1994), {\it Nature} {\bf 372}, 652.
\bibitem[]{ja00} Jain, P., {\it et al.} (2000), {\it Physics Letters B}
{\bf 484}, 267.
\bibitem[]{jo01} Jorstad, S.G. \etal (2001), \apj {\it Suppl.} {\bf 134}, 181.
\bibitem[]{jo00a} Jones, F.C. (2000), paper presented at the {\it OWL Workshop, 
2000}.
\bibitem[]{jo00} Jones, T.W. (2000), in {\it Proc. Seventh Taipei Workshop
on Astrophysics}, e-print astro-ph/0012483.
\bibitem[]{ka21} Kaluza, T. (1921) {\it Sitzungsberichten der Preussen Akademie
f\"{u}r Wissenschaften, Berlin (Math. Phys.)} {\bf K1}, 966.
\bibitem[]{ka86} Kazanas, D. and Ellison, D.C. (1986), \apj {\bf 304}, 178.
\bibitem[]{ki76} Kibble, T.W.B. (1976), {\it J. of Phys. A} {\bf9}, 1387.
\bibitem[]{ki99}T. Kifune (1999), \apj {\bf 518}, L21.
\bibitem[]{ki95} Kifune, T. \etal\ (1995), \apj, {\bf 438}, L91. 
\bibitem[]{ki01} Kiss, Cs. \etal\ (2001), \aa\ {\bf 379}, 1161.
\bibitem[]{kl26} Klein, O. (1926), {\it Zeitschrift f\"{u}r Physik} {\bf 37}, 
895.
\bibitem[]{ko90} Kolb, E.W. and Turner, M.S. (1990), {\it The Early Universe},
(Reading: Addison Wesley).
\bibitem[]{kr00} Krawczynski\ \etal (2000), \aa\ {\bf 353}, 97.
\bibitem[]{kr99} Krennrich, F. {\it et al.}, (1999), \apj {\bf 511}, 149.
\bibitem[]{kr02} Krennrich, F. {\it et al.}, (2002), \apj {\bf 575}, L9.
\bibitem[]{kr98} Krizmanic, J.F. Ormes, J.F. and Streitmatter, R.E. (1998),  
{\it Workshop on  Observing Giant Cosmic Ray Air Showers from Space, AIP 
CP-433} (New York: American Institute of Physics).
\bibitem[]{kro99} Krolik, J. (1999), \apj {\bf 515}, L73.
\bibitem[]{ku98} Kulkarni, S.R. \etal\ (1998), {\it Nature} {\bf 393}, 35.
\bibitem[]{kuz98} Kuz'min, V.A., and Rubakov, V.A. (1998), {\it Phys. 
Atom Nucl.} {\bf 61}, 1028.
\bibitem[]{la00} Lagache, G. \etal (2000), \aa\ {\bf 354}, 247.
\bibitem[]{la53} Landau, L.D. and Pomeranchuk, I.J. (1953), {\it Doklady
Acad. Nauk SSSR} {\bf 92}, 535.
\bibitem[]{le95} Learned, J.G. and Pakvasa, S. (1995), {\it Astropart. Phys.}
{\bf 3}, 267.
\bibitem[]{li96} Lilly, S.J. \etal\ (1996), \apj\ {\bf 460}, L1.
\bibitem[]{li63} Linsley, J. (1963), {\prl},{\bf 10}, 146.
\bibitem[]{liv98} Livio, M. \etal\ (1998), in {\it Proc. 4th Huntsville Symp. 
on Gamma Ray Bursts, AIP CP-428}, ed. C.A. Meegan\ \etal\ (New York: American 
Institute of Physics) p. 483.
\bibitem[]{ma94} Ma, C.-P. and Bertschinger, E. (1994), \apj {\bf 434}, L5.
\bibitem[]{mad96} Madau, P. and Schull (1996), \apj\ {\bf 457}, 551.
\bibitem[]{ma01} Madau, P., and Pozzetti, L., (2001), \mnras {\bf 312}, L9.
\bibitem[]{ms98} Malkan, M.A. and Stecker F.W. (1998), \apj {\bf 496}, 13.
\bibitem[]{ms01} Malkan, M.A. and Stecker F.W. (2001), \apj {\bf 555}, 641.
\bibitem[]{ma93} Mannheim, K. (1993), \aa\ {\bf 269}, 67.
\bibitem[]{ma98} Mao, S. and Mo, H.J. (1998), {\it Astron. Astrophys.} 
{\bf 339}, L1.
\bibitem[]{ma96} Mastichiadas, A. and de Jager, O.C. (1996), \aa\ {\bf 305}, 
L5.
\bibitem[]{ma94} Mather, J. {\it et al.} (1994), \apj {\bf 420}, 439.
\bibitem[]{me01} M\'{e}sz\'{a}ros, P. and Waxman, E. (2001), \prl {\bf 87}, 
171102.
\bibitem[]{mi56} Migdal, A.B. (1956), {\it Phys. Rev.} {\bf 103}, 1811.
\bibitem[]{mu00} Muraishi, H.\ \etal\ (2000), \aa\, {\bf 354}, 57.
\bibitem[]{na00} Nagano, M. and Watson, A.A. (2000), {\it Rev. Mod. 
Phys.} {\bf 72}, 689.
\bibitem[]{ni80} Nishimura, J. \etal (1980), Astrophys. J. {\bf 238}, 394.
\bibitem[]{no02} Norris, J. (2002), \apj , in press, e-print astro-ph/0201503.
\bibitem[]{nu99} Nussinov, S. and Schrock, R. (1999), {\it Phys. Rev. D}
{\bf 59}, 105002. 
\bibitem[]{ol00} Olinto, A (2000), in {\it Observing Ultrahigh Energy
Cosmic Rays from Space and Earth, AIP CP-566} (New York: American Institute 
of Physics), pg. 99, e-print astro-ph/0011106.
\bibitem[]{pe65} Penzias, A.A. and Wilson, R.W. (1965), \apj {\bf 142}, 419.
\bibitem[]{pe00} Petry, D. {\it et al.} (2000), \apj {\bf 536}, 742.
\bibitem[]{pe02} Petry, D. \etal , (2002), \epr 0207506.
\bibitem[]{pe94} Pettini M.\ \etal\ (1994), \apj\ {\bf 426}, 79.
\bibitem[]{pi98} Pian, E. {\it et al.} (1998), \apj\ {\bf 492}, L17
\bibitem[]{po96} Pohl, M. \etal\ (1996), \aa\ {\bf 307}, 57.
\bibitem[]{pr96} Protheroe, R.J. and Biermann, P.L. (1996), {\it Astroparticle 
Physics} {\bf 6}, 45.
\bibitem[]{pu76} Puget, J.-L., Stecker, F.W. and Bredekamp, J. (1976), \apj 
{\bf 205}, 638.
\bibitem[]{pu96} Puget, J.-L. \etal (1996), \aa\ {\bf 308}, L5.
\bibitem[]{pu99} P\"{u}hlhofer, G. \etal (1999), \xxviicrc , {\bf 3}, 492 .
\bibitem[]{pu92} Punch, M. \etal\ (1992), {\it Nature} {\bf 358}, 477.
\bibitem[]{qu96} Quinn, J. \etal\ (1996), \apj\ {\bf 456}, L83.
\bibitem[]{ra99} Randall, L. and Sundrum, R. (1999), \prl\ {\bf 83}, 3370.
\bibitem[]{re01} Renault, C.\ \etal (2001), \aa\ {\bf 371}, 771.
\bibitem[]{ro88} Robinett, R.W. (1988), {\it Phys. Rev. D} {\bf 37}, 84.
\bibitem[]{ro} Rowan-Robinson, M. (2001), \apj {\bf 549}, 745.
\bibitem[]{sa68} Salam, A. (1968), in {\it Elementary Particle Theory (Proc. 
8th Nobel Symp.)}, ed. N. Svartholm (Stockholm: Almqvist and Wiksells Vorlag), 
pg. 367. 
\bibitem[]{ss98} Salamon, M.H. and Stecker, F.W. (1998), \apj {\bf 493}, 547 
(SS98).
\bibitem[]{sa01} Sarkar, S. and Toldr\`{a}, R. (2002), {\it Nucl. Phys. B}
{\bf 621}, 495.
\bibitem[]{sas01} Sasaki, N. \etal (2001), {\it Proc. 27th Intl. Cosmic Ray
Conf., Hamburg} {\bf 1}, 333.
\bibitem[]{sa72} Sato, H. and Tati, T. (1972), {\it Progr. Theor. 
Phys.} {\bf 47}, 1788.
\bibitem[]{sc99} Schmidt, M. (1999), \apj\ {\bf 523}, L117.
\bibitem[]{sc89} Schaefer, R.K., Shafi, Q. and Stecker, F.W. (1989), \apj\
{\bf 347}, 575.
\bibitem[]{sc02} Scully, S.T. and Stecker, F.W. (2002) {\it Astroparticle 
Physics} {\bf 16}, 271.
\bibitem[]{sh84} Shafi, Q. and Stecker, F.W. (1984), \prl {\bf 53}, 1292.
\bibitem[]{sh02} Shinozaki, {\it et al.} (2002), \apj {\bf 571}, L117.
\bibitem[]{sh53} Shklovskii, I. S. (1953), {\it Doklady Akad. Nauk SSSR} 
{\bf 91}, 475.
\bibitem[]{si98} Sigl, G. \etal\ (1998), {\it Phys. Rev. D} {\bf 59}, 043504.
\bibitem[]{sm92} Smoot, G. {\it et al.} (1992), \apj {\bf 396}, L1.
\bibitem[]{sm02} Smy, M.B.,  (2002), e-print hep-ex/0208004.
\bibitem[]{so98} Sokolsky, P. (1998), in {\it Workshop on Observing Giant 
Cosmic Ray Air Showers from Space}, ed. J.F. Krizmanic, J.F. Ormes and R.E. 
Streitmatter, (New York: American Institute of Physics) p. 65. 
\bibitem[]{sr98} Sreekumar, P. \etal\ (1998), \apj\ {\bf 494}, 523.
\bibitem[]{stan98} Stanev, T., and Franceschini, A., (1998), \apj, {\bf 494}, 
L159.
\bibitem[]{st95} Stanev, T. {\it et al.} (1995), \prl {\bf 75}, 3056.
\bibitem[]{st68} Stecker, F.W. (1968), \prl {\bf 21}, 1016. 
\bibitem[]{st69} Stecker, F.W. (1969), \apj {\bf 157}, 507
\bibitem[]{st73} Stecker, F.W. (1973), {\it Astrophys. Space Sci.}
{\bf 20}, 47.
\bibitem[]{st79} Stecker, F.W. (1979), \apj {\bf 228}, 919.  
\bibitem[]{st89} Stecker, F.W. (1989), {\it Nature} {\bf 342}, 401.
\bibitem[]{st99} Stecker, F.W. (1999), {\it Astropart. Phys.} {\bf 11}, 83.
\bibitem[]{st00} Stecker, F.W. (2000), {\it Astropart. Phys.} {\bf 14}, 207.
\bibitem[]{st01} Stecker, F.W. (2001),  in
{\it The Extragalactic Infrared Background and its Cosmological 
Implications, IAU Symposium no. 204}, ed. M. Harwit and M.G. Hauser,
p. 135.
\bibitem[]{sd93} Stecker, F.W. and De Jager, O.C. (1993), \apj\ {\bf 415}, L71.
\bibitem[]{sd96} Stecker, F.W. and De Jager, O.C. (1996), {\it Space Sci. 
Rev.} {\bf 75}, 413.
\bibitem[]{st97} Stecker, F.W. and de Jager, O.C. (1997), in {\it Towards a 
Major Atmospheric Cerenkov Detector V, Proc.
Kruger Natl. Park Workshop on TeV Gamma-Ray Astrophysics, Berg-en-Dal,
South Africa}, ed. O.C. de Jager, (Potchefstoom: Wesprint), p. 39, e-print 
astro-ph/9710145.
\bibitem[]{sd97} Stecker, F.W. and De Jager, O.C. (1997), \apj\ {\bf 476}, 712.
\bibitem[]{st98} Stecker, F.W. and de Jager, O.C. (1998), \aa\ {\bf 334}, L85.
\bibitem[]{st92} Stecker, F.W. de Jager, O.C. and Salamon, M.H. (1992), \apj\
{\bf 390}, L49.
\bibitem[]{st96a} Stecker, F.W. De Jager, O.C. and Salamon, M.H. (1996), 
\apj\ {\bf473}, L75.
\bibitem[]{sg01} Stecker, F.W. and Glashow, S.L. (2001), {\it Astropart.
Phys.} {\bf 16}, 97.
\bibitem[]{ss96} Stecker, F.W. and Salamon, M.H. (1996a), \apj\ {\bf 464}, 600.
\bibitem[]{st96b} Stecker, F.W. and Salamon, M.H. (1996b), {\it Space Sci. 
Rev.} {\bf 75}, 341.
\bibitem[]{st97} Stecker, F.W. and Salamon, M.H. (1997), \xxvicrc\ {\bf 3}, 
317.
\bibitem[]{st99} Stecker, F.W. and Salamon, M.H. (1999), \apj {\bf 512}, 521.
\bibitem[]{st91} Stecker, F.W. {\it et al.} (1991), \prl {\bf 66}, 2697, E:
{\bf 69}, 2738.  
\bibitem[]{stei98} Steidel, C.C. \etal\ (1999), {\it Proc Natl. Acad. Sci. USA}
{\bf 96}, 4232.
\bibitem[]{ta98} Takeda, M. {\it et al.} (1998), \prl {\bf 81}, 1163.
\bibitem[]{ta99} Takeda, M. {\it et al.} (1999) \apj {\bf 522}, 225.
\bibitem[]{tan99} Tan, J.C., Silk, J. and Balland, C. (1999), \apj {\bf522}, 579.
\bibitem[]{ta98a} Tanimori, T. \etal (1998a), \apj {\bf 497}, L25.
\bibitem[]{ta98b} Tanimori, T. \etal (1998b), \apj {\bf 492}, L33.
\bibitem[]{tr79} Tremaine, S. and Gunn, J.E. (1979), \prl {\bf 42}, 407.
\bibitem[]{ty01} Tyler, C., Olinto, A.V. and Sigl, G. (2001), {\it Phys. 
Rev. D} {\bf 63}, 055001.
\bibitem[]{vi95} Vietri, M. (1995) \apj {\bf 453}, 883.
\bibitem[]{wa95} Waxman, E. (1995), \prl {\bf 75}, 386.
\bibitem[]{wa97} Waxman, E. and Bahcall, J. (1997), {\it Phys. Rev. Letters} 
{\bf78}, 2292.
\bibitem[]{wd72} Wdowczyk, J., Tkaczyk, T. and Wolfendale, A.W. (1972),
{\it J. Phys. A} {\bf 5}, 1419.
\bibitem[]{we89} Weekes, T.C. \etal (1989), \apj {\bf 342}, 379.
\bibitem[]{we82} Weiler, T.J. (1982), \prl {\bf 49}, 234.
\bibitem[]{we99a} Weiler, T.J. (1999), {\it Astropart. Phys.} {\bf 11}, 303.
\bibitem[]{we67} Weinberg, S. (1967), \prl {\bf 19}, 1264.
\bibitem[]{we76} Weinberg, S. (1979), \prl {\bf 42}, 850. 
\bibitem[]{we99} Weinheimer, C., {\it et al.} (1999), {\it Phys. Lett. B}
{\bf 460}, 219.
\bibitem[]{wi02} Wischnewski, R. (2002), \epr 0204268.
\bibitem[]{wr92} Wright, E.L. (1992), \apj {\bf 396}, L13.
\bibitem[]{wr00} Wright, E.L. and Reese, E.D. (2000), \apj, 545, 43
\bibitem[]{yo98} Yoshida, S. Sigl, G. and Lee S. (1998), {\it Phys. Rev. 
Letters} {\bf 81}, 5505.
\bibitem[]{yo97} Yoshikoshi, T. \etal (1997), \apj {\bf 487}, L65.
\bibitem[]{xu00} Xu, C. (2000), \apj {\bf541}, 134.
\bibitem[]{za01} Zas, E. (2001), in {\it Radio Detection of High Energy
Particles, AIP CP-579}, ed. D. Saltzberg and P. Gorham, 
(New York: American Institute of Physics), p. 286, \epr 0103371.
\bibitem[]{za66} Zatsepin, G.T. and Kuz'min, V.A. (1966), {\it Zh. Esks. Teor. 
Fiz., Pis'ma Red.} {\bf 4}, 144.


\end{thebibliography}
\end{document}